\documentclass[final,5p,times,twocolumn,number,sort&compress]{elsarticle}

\usepackage{amssymb}
\usepackage{lipsum}
\usepackage{amsmath}
\usepackage{graphicx}

\usepackage{xcolor}

\journal{Physics Letters B}

\begin{document}

\begin{frontmatter}



\title{Theoretical Consistency and Observational Constraints on Entropic Cosmology}

\affiliation[affiliation1]{organization={School of Physics Science and Technology, Xinjiang University},
            city={Urumqi},
            postcode={830017}, 
            country={China}}

\affiliation[affiliation2]{organization={School of Physics and Electronics, Henan University},
            city={Kaifeng},
            postcode={475004}, 
            country={China}}

\affiliation[affiliation3]{organization={Xinjiang Astronomical Observatory, Chinese Academy of Sciences},
            addressline={150 Science 1-Street}, 
            city={Urumqi},
            postcode={830011}, 
            country={China}}

\affiliation[affiliation4]{organization={School of Astronomy and Space Science, University of Chinese Academy of Sciences},
            addressline={No.19A Yuquan Road}, 
            city={Beijing},
            postcode={100049}, 
            country={China}}

\affiliation[affiliation5]{organization={School of Physics and Astronomy, China West Normal University},
            city={Nanchong},
            postcode={637002}, 
            country={China}}

\author[affiliation1,affiliation2]{Yunliang Ren}
\author[affiliation2]{Xiaofeng Yang\corref{cor1}}
\author[affiliation3,affiliation4]{Xuwei Zhang}
\author[affiliation1,affiliation2]{Shuangnan Chen}
\author[affiliation5,affiliation2]{Yangjun Shi}
\author[affiliation1]{Hoernisa Iminniyaz\corref{cor2}}

\cortext[cor1]{Corresponding author. Email: xfyang@henu.edu.cn}
\cortext[cor2]{Corresponding author. Email: wrns@xju.edu.cn}

\begin{abstract}

Recent DESI DR2 results suggest that dark energy may evolve with time, which motivates a re-examination of entropic cosmology.
In this work, we study three representative entropic models, denoted as Models A, B, and C, 
corresponding to the driving terms $H^2$, $H^2+\dot H$, and $H^2+\dot H+H^4$, respectively.
We establish a general framework for entropic cosmology and examine these models through early universe consistency, 
the Trans-Planckian Censorship Conjecture (TCC), and the realization of Quintom-B crossing.
We constrain these models using the Pantheon+ sample, DESI DR2 BAO measurements, 
and Planck 2018 CMB distance priors, and compare their fits to the data.
For Model C, we adopt a first-order perturbative treatment in the full data analysis 
to maintain numerical stability in the presence of the higher-order term.
We find that Model A is the simplest among the three models, 
but it is difficult to make it compatible with both early universe consistency and the TCC condition.
Model B, which includes a derivative correction term, 
shows a small difference in the fit to observational data compared to Model A, 
and is more consistent with the theoretical constraints.
Model C includes more correction terms, but does not provide a better fit to the data.
The parameter region favored by observations does not fully overlap with that preferred by theoretical consistency, 
which suggests that entropic cosmology may require further development to achieve a self-consistent description across different cosmic epochs.

\end{abstract}


\end{frontmatter}

\section{Introduction}

Dark energy is one of the most challenging topics in contemporary cosmological research, 
and its physical properties have not yet been fully revealed. 
In 1998 and 1999, 
two independent supernova observation teams discovered the accelerated expansion of the Universe, 
providing compelling evidence 
that the late-time cosmic expansion is accelerating \cite{acceleration1, acceleration2, acceleration3, acceleration4}. 
In standard cosmology, this acceleration is commonly attributed to dark energy. 
Although many theoretical models have been proposed, 
no one can fully explain the physical properties of dark energy or agree with all observations. 
The $\Lambda$ cold dark matter ($\Lambda$CDM) model is currently the most popular and simplest model for the acceleration of the universe, 
and it is known as the standard cosmological model. 
In $\Lambda$CDM, dark energy is the main component driving the accelerated expansion of the universe, 
accounting for about 70$\%$ of the total energy density, and exhibits negative pressure characteristics. 
The $\Lambda$CDM model has been used as a standard reference for cosmological observations and data analysis for the past few decades. 
Although the model has been successful in observation, it still faces some theoretical challenges, 
such as cosmological constant problems and coincidence problems \cite{LCDM-problem1, LCDM-problem2}. 
This has motivated the exploration of dynamical dark energy models beyond the cosmological constant \cite{dynamical1, dynamical2, 
dynamical3, dynamical4, dynamical5, dynamical6, dynamical7, dynamical8, dynamical9, dynamical10, dynamical11, dynamical12, dynamical13}.

The close connections among gravity, thermodynamics, 
and the holographic principle have provided new perspectives in gravitational physics and cosmology. 
In Verlinde's entropic-gravity proposal, gravity is not regarded as a fundamental interaction, 
but as an emergent phenomenon associated with changes in entropy and information \cite{Verlinde,Verlinde2017}. 
In this proposal, the displacement of matter relative to a holographic screen changes the entropy 
and information associated with the screen, 
thereby giving rise to an entropic force that can reproduce gravitational behavior. 
Entropic cosmology, proposed by Easson, Frampton, and Smoot \cite{Easson,EFDE7}, 
adopts a different approach. It retains general relativity as the underlying gravitational theory, 
while additional entropic-force terms are introduced into the cosmological field equations. 
These terms are motivated by the holographic principle, horizon thermodynamics, 
and boundary contributions to the Einstein-Hilbert action, 
and can provide an alternative mechanism for cosmic acceleration 
without invoking a conventional dark-energy component \cite{EFDE1, EFDE2, EFDE3, EFDE4, EFDE5, EFDE6, EFDE7, 
EFDE8, EFDE9, EFDE10, EFDE11, EFDE12, EFDE13}. 
It is important to clarify that, 
although entropic cosmology and Verlinde's entropic-gravity proposal both involve concepts related to the holographic principle, 
entropy, and horizon thermodynamics, 
they are conceptually distinct theories based on different physical assumptions \cite{EFDE4}.

Recent results from the Dark Energy Spectroscopic Instrument (DESI) 
suggest that dark energy may evolve with time \cite{DESI1, DESI2, DESI3, DESI4, DESI5}, 
which motivates the study of models beyond the cosmological constant. 
In this context, entropic cosmology has attracted renewed interest, 
since it naturally introduces dynamical modifications to the cosmic expansion history \cite{latest-EFDE1, latest-EFDE2, 
latest-EFDE3, latest-EFDE4, latest-EFDE5, latest-EFDE6}. 
Although no fundamental dark-energy component is introduced in entropic cosmology, 
the modified background equations can be rewritten in an effective-fluid form. 
The effective energy density and equation-of-state parameter employed 
in this work are introduced to characterize the expansion history 
and facilitate comparison with cosmological observations. 
Within this effective description, 
the cosmological evolution of the models can be described in terms of quantities 
such as the effective equation-of-state parameter and the deceleration parameter, 
and can be compared directly with current observations, including Type Ia supernovae, 
Cosmic Microwave Background (CMB) distance priors, and DESI Baryon Acoustic Oscillation (BAO) measurements.

Within this effective description, 
we consider three previously studied classes of entropic driving terms, 
denoted as Models A, B, and C, corresponding to $H^2$, $H^2+\dot{H}$, 
and $H^2+\dot{H}+H^4$, respectively. 
The inclusion of the $\dot H$ and $H^4$ contributions provides a direct way to examine, 
within the same set of cosmological equations, how derivative and higher-order entropic terms 
modify the cosmic expansion history relative to the $H^2$ case. 
Previous analyses have examined both the expansion history 
and the growth of linear matter perturbations in related entropic models 
\cite{EFDE12,EFDE8}. 
Related developments have considered non-adiabatic-like expansion in entropic cosmology 
\cite{EFDE13}, 
entropic-force models based on a generalized black-hole entropy 
\cite{EFDE6}, 
and a phenomenological dissipative model with two constant terms 
\cite{KomatsuKimura2015}. 
Against this broader theoretical background, we examine the theoretical consistency 
of the expansion histories of the three models and assess their observational viability 
using current cosmological data. 
The corresponding structure-growth behavior depends on additional perturbative relations 
beyond the background equations and therefore requires a dedicated perturbation analysis. 
Starting from a general parameterization of entropic cosmology, 
we derive the general expressions for the effective equation of state, 
and obtain the corresponding effective equation of state, deceleration parameter, 
and Hubble expansion rate $E(z)$ for each model. 
We examine the three models separately under the early-universe conditions, 
the Trans-Planckian Censorship Conjecture (TCC), 
and the Quintom-B evolution hinted at by recent DESI analyses. 
We constrain the models using the Pantheon+ supernova sample, DESI DR2 BAO data, 
and CMB distance priors from Planck 2018. 
In particular, for Model C, the coefficient of the $H^4$ term is extremely small, 
which makes stable parameter estimation difficult when the CMB distance prior is included. 
We adopt a first-order perturbative approximation only for the full data combination 
to control numerical uncertainties at high redshift. For the late-time analysis using Pantheon+ and DESI alone, 
the exact solution is used. 
The structure of this paper is as follows: 
Section 2 presents the general framework and the detailed derivation of the three models, 
including their effective equations of state and evolution equations; 
Section 3 discusses the constraints from the early universe, TCC, and Quintom-B evolution; 
Section 4 presents the observational datasets, the fitting results, 
and the $\chi^2$ and Akaike Information Criterion(AIC) analyses; 
Section 5 summarizes the main conclusions of this work.

\section{Entropic Cosmology}
\label{sec:models}

In entropic cosmology, the accelerated expansion of the universe is attributed to thermodynamic effects associated with the cosmic horizon. 
The horizon is assumed to be a thermodynamic boundary associated with an entropy $S$ and a temperature $T$, 
as motivated by the holographic principle. 
This thermodynamic description is motivated by the Bekenstein entropy-area relation 
and Hawking's black-hole temperature \cite{Bekenstein1973,Hawking1974}.
The entropic force is written as:
\begin{equation}\label{eq:entropy force}
F = -T \frac{dS}{dr_H},
\end{equation}
where $r_H$ is the radius of the Hubble horizon. 
Using the Bekenstein-Hawking entropy $S = k_B c^3 A / (4 G \hbar)$ and the horizon temperature $T = \hbar H / (2\pi k_B)$, 
the corresponding entropic-force pressure on the horizon can be expressed as:
\begin{equation}\label{eq:effective pressure}
P_F = \frac{F}{A} = -\frac{T}{A} \frac{dS}{dr_H}.
\end{equation}
Using $A = 4\pi r_H^2$, we find
\begin{equation}\label{eq:dS/dr}
\frac{dS}{dr_H} = \frac{k_B c^3}{4 G \hbar} \frac{d}{dr_H}(4\pi r_H^2) = \frac{2\pi k_B c^3 r_H}{G \hbar}.
\end{equation}
Substituting the expressions for $T$ and $dS/dr_H$ into Eq.(\ref{eq:effective pressure}), we obtain
\begin{equation}\label{eq:pressure}
\begin{aligned}
P_F &= -\frac{1}{4\pi r_H^2}
\left( \frac{\hbar H}{2\pi k_B} \right)
\left( \frac{2\pi k_B c^3 r_H}{G \hbar} \right) \\
&= -\frac{c^3 H}{4\pi G r_H}
= -\frac{c^2 H^2}{4\pi G}.
\end{aligned}
\end{equation}
Here, the negative pressure term ($P_F \propto -H^2$) contributes to cosmic acceleration. 
More general entropy-area relations or higher-order corrections can lead to additional terms involving $\dot{H}$ or higher powers of $H$. 
We parameterize these corrections as an extra driving term $g(t)$ in the cosmological equations.

\subsection{General Dynamical Equations}

Entropic effects are described by two functions, $f(t)$ and $g(t)$, 
which modify the Friedmann equation and the acceleration equation, 
respectively. We set $c=1$. The modified equations are
\begin{equation}
H^2=\frac{8\pi G}{3}\left(\rho_m+\rho_r\right)+f(t),
\label{eq:Friedmann_fg}
\end{equation}
and
\begin{equation}
\frac{\ddot a}{a}=-\frac{4\pi G}{3}\left(\rho_m+2\rho_r\right)+g(t).
\label{eq:acceleration_fg}
\end{equation}

In this work, the matter and radiation components are assumed to be separately conserved:
\begin{equation}
\dot{\rho}_m+3H\rho_m=0,
\label{eq:matter_cons}
\end{equation}
\begin{equation}
\dot{\rho}_r+4H\rho_r=0.
\label{eq:radiation_cons}
\end{equation}
With this assumption, $f(t)$ and $g(t)$ are not independent. 
Taking the time derivative of Eq.(\ref{eq:Friedmann_fg}) and combining it with Eq.(\ref{eq:acceleration_fg}), we obtain
\begin{equation}
\frac{d f(t)}{dt}+2Hf(t)=2Hg(t).
\label{eq:fg_consistency}
\end{equation}
This relation implies that, once the standard conservation laws for matter and radiation are imposed, the function $f(t)$ is fixed once $g(t)$ is given.

For the purpose of describing the expansion history and comparing the models with cosmological observations, 
we introduce an effective dark-energy density associated with the contribution $f(t)$ and define
\begin{equation}
\rho_{\rm DE} = \frac{3}{8\pi G}f(t),
\label{eq:rhoDE_def}
\end{equation}
so that
\begin{equation}\label{Friedmann}
H^2=\frac{8\pi G}{3}\left(\rho_m+\rho_r+\rho_{\rm DE}\right).
\end{equation}
This effective description does not imply the presence of a separate fundamental dark-energy component, 
but provides a convenient form for confronting the models with cosmological observations. 
Entropic models without a corresponding contribution to the Friedmann equation 
generally have difficulty reproducing a realistic matter density parameter $\Omega_m$ \cite{latest-EFDE5}.

Using the definition of $\rho_{\rm DE}$ in Eq.(\ref{eq:rhoDE_def}), Eq.(\ref{eq:fg_consistency}) becomes
\begin{equation}
\dot{\rho}_{\rm DE}+2H\rho_{\rm DE}=\frac{3H}{4\pi G}g(t),
\label{eq:rhoDE_consistency}
\end{equation}
so that, once matter and radiation obey the standard conservation laws, the evolution of $\rho_{\rm DE}$ is fixed by the entropic driving term $g(t)$.

The acceleration equation is modified by the entropic driving term and can be written as
\begin{equation}\label{acceleration equation}
\frac{\ddot a}{a}=-\frac{4\pi G}{3}\left(\rho_m+2\rho_r\right)+g(t),
\end{equation}
where $g(t)$ represents the entropic contribution to the cosmic acceleration. 
We next derive the total effective equation of state, the effective equation of state of the entropic dark energy component, 
and the deceleration parameter.

We define the total effective equation-of-state parameter, $w_{\rm eff}$, 
for the combined matter, radiation, and entropic contributions. Using
\begin{equation}
\frac{\ddot a}{a}=-\frac{H^2}{2}(1+3w_{\rm eff}),
\label{eq:acc_weff}
\end{equation}
together with Eq.(\ref{acceleration equation}), and introducing the dimensionless density parameters
\begin{equation}
\Omega_i(z) = \frac{8\pi G\rho_i}{3H^2},
\label{eq:Omega_def}
\end{equation}
we find
\begin{equation}
w_{\rm eff}(z)=\frac{\Omega_m(z)+2\Omega_r(z)-1}{3}-\frac{2g(t)}{3H^2(z)}.
\label{eq:weff_general}
\end{equation}

We also define an effective equation-of-state parameter for the entropic dark energy. From
\begin{equation}
w_{\rm eff}=\Omega_m w_m+\Omega_r w_r+\Omega_{\rm DE}w_{\rm DE},
\label{eq:weff_decomp}
\end{equation}
with $w_m=0$ for pressureless matter and $w_r=1/3$ for radiation, we obtain
\begin{equation}
w_{\rm DE}(z)=\frac{\Omega_m(z)+\Omega_r(z)-1-\frac{2g(t)}{H^2(z)}}{3\Omega_{\rm DE}(z)}.
\label{eq:wDE_general}
\end{equation}
For a spatially flat universe, the density parameters satisfy the relation
\begin{equation}
\Omega_{\rm DE}(z)=1-\Omega_m(z)-\Omega_r(z).
\label{eq:closure}
\end{equation}

Finally, substituting Eq.(\ref{acceleration equation}) into the definition
\begin{equation}
q = -\frac{\ddot a}{aH^2},
\label{eq:q_def}
\end{equation}
we obtain
\begin{equation}
q(z)=\frac{1}{2}\Omega_m(z)+\Omega_r(z)-\frac{g(t)}{H^2(z)}.
\label{eq:q_general}
\end{equation}

After introducing $\rho_{\rm DE}$, the effect of entropic corrections is reflected in the form of $g(t)$. 
According to the form of $g(t)$, we divide the models considered in this work into three classes: Model A, Model B, and Model C.

The three representative forms considered below are motivated by developments 
in the entropic-cosmology literature. 
The $H^2$ contribution follows from the horizon entropy-temperature construction 
\cite{Easson}. 
The surface-term treatment of Easson, Frampton, and Smoot further introduced 
contributions proportional to both $H^2$ and $\dot{H}$ 
\cite{Easson,EFDE7}. 
In the same framework, logarithmic quantum corrections to the Bekenstein-Hawking 
entropy were shown to generate an additional $H^4$ term 
\cite{EFDE7}. 
A general parameterization containing $H^2$, $\dot{H}$, and $H^4$ contributions 
was subsequently examined in Ref.~\cite{EFDE3}, where the higher-order terms were 
discussed in connection with quantum corrections to the entropy-area relation.

Boundary contributions to the gravitational action provide an additional 
theoretical perspective on entropic acceleration. 
For spacetimes with boundaries, the Einstein-Hilbert action is supplemented by the 
Gibbons-Hawking-York boundary term \cite{York1972,GibbonsHawking1977}. 
This connection has recently been emphasized in the General Relativistic Entropic 
Acceleration (GREA) framework, which provides a covariant formalism for 
out-of-equilibrium dynamics in general relativity and relates cosmic acceleration 
to the growth of entropy associated with cosmic and black-hole horizons 
\cite{GarciaBellido2024a,GarciaBellido2024b}. 
In GREA, the entropic contribution is associated with the growth of horizon entropy 
and the corresponding GHY boundary term in the gravitational action, 
while the present work examines the theoretical consistency and observational viability 
of the previously proposed parameterized forms of $g(t)$.

Previous studies have examined the cosmological evolution 
of related entropic models. 
For the entropic formulations containing $H^2$ and $\dot H$ discussed in 
Ref.~\cite{EFDE12}, 
an additive constant in the Friedmann equations was found to play an important role 
in realizing the transition from decelerated to accelerated expansion. 
The subsequent analysis of Ref.~\cite{EFDE8} extended the analysis 
to the growth of linear matter perturbations, providing an additional test of these models. 
These studies considered both the expansion history and the growth of cosmological perturbations 
in assessing entropic cosmological models. 
In the present formulation, matter and radiation satisfy the standard conservation equations, 
which relate the two functions $f(t)$ and $g(t)$ through 
Eq.~(\ref{eq:fg_consistency}). 
The resulting cosmological evolution is examined through the Hubble expansion rate, 
the deceleration parameter, and the early-universe consistency conditions. 
The perturbative evolution of the entropic contribution requires additional relations 
beyond the background equations. 
Accordingly, the implications of the present parameterizations for structure growth 
require a dedicated perturbation analysis.

More recently, Gohar and Salzano reassessed the thermodynamic foundations
of entropic cosmology and showed that, when the entropy-temperature relation is
implemented consistently, the modified entropic-force models considered in their analysis
become thermodynamically equivalent to the original Bekenstein-Hawking formulation
and retain the same phenomenological limitations unless more fundamental assumptions
are modified \cite{latest-EFDE4}.
In the present work, the entropic corrections are parameterized directly through the
driving term $g(t)$ in the cosmological equations. Models A, B, and C are specified by the
$H^2$, $H^2+\dot H$, and $H^2+\dot H+H^4$ forms of this term, respectively.
The analysis tests the theoretical conditions and observational viability
of these previously proposed parameterizations.

\subsection{Model A}

Model A has the simplest form of the three models, with
\begin{equation}
g(t)=\beta_1 H^2.
\label{eq:gA}
\end{equation}
This corresponds to an entropic-force term of the $H^2$ form.  
The Bekenstein-Hawking entropy is a representative example, 
but the same $H^2$ structure also appears in more general thermodynamically consistent entropic models \cite{Easson, latest-EFDE5, latest-EFDE6}.

Substituting Eq.(\ref{eq:gA}) into Eq.(\ref{eq:q_general}), we obtain the deceleration parameter
\begin{equation}
q(z)=\frac{1}{2}\Omega_m(z)+\Omega_r(z)-\beta_1,
\label{eq:qA}
\end{equation}
while substituting Eq.~(\ref{eq:gA}) into Eq.~(\ref{eq:wDE_general}) together with Eq.~(\ref{eq:closure}), we obtain the effective equation of state of the entropic dark energy,
\begin{equation}
w_{\rm DE}(z)=\frac{\Omega_m(z)+\Omega_r(z)-1-2\beta_1}{3\left[1-\Omega_m(z)-\Omega_r(z)\right]}.
\label{eq:wDEA}
\end{equation}
In Model A, the dependence of $q(z)$ and $w_{\rm DE}(z)$ on the entropic term is governed by $\beta_1$.

To derive the dimensionless Hubble rate, we define
\begin{equation}
E^2(z) = \frac{H^2(z)}{H_0^2},
\label{eq:E2_def_A}
\end{equation}
and rewrite the definition of the deceleration parameter as
\begin{equation}
q(z)=-1+\frac{1+z}{2}\frac{d\ln E^2}{dz}.
\label{eq:qE_relation_A}
\end{equation}
Combining Eq.(\ref{eq:qA}) with Eq.(\ref{eq:qE_relation_A}), and using
\begin{equation}
\Omega_m(z)=\frac{\Omega_{m0}(1+z)^3}{E^2(z)}, \qquad
\Omega_r(z)=\frac{\Omega_{r0}(1+z)^4}{E^2(z)},
\label{eq:OmegaA_sub}
\end{equation}
we obtain a first-order linear differential equation for $E^2(z)$. Solving it with the boundary condition $E^2(0)=1$, we find
\begin{equation}
\begin{aligned}
E^2(z)&=\frac{\Omega_{m0}}{1+2\beta_1}(1+z)^3
+\frac{\Omega_{r0}}{1+\beta_1}(1+z)^4 \\[3pt]
&\quad+\left(1-\frac{\Omega_{m0}}{1+2\beta_1}
-\frac{\Omega_{r0}}{1+\beta_1}\right)(1+z)^{2(1-\beta_1)},
\end{aligned}
\label{eq:E2A}
\end{equation}
where $\Omega_{m0}$ and $\Omega_{r0}$ denote the matter and radiation density parameters at $z=0$, respectively. 
Eq.(\ref{eq:E2A}) shows that the parameter $\beta_1$ modifies the coefficients of the matter and radiation terms 
and introduces an additional contribution proportional to $(1+z)^{2(1-\beta_1)}$. 
This solution is valid for $\beta_1\neq -1$ and $\beta_1\neq -1/2$.

\subsection{Model B}

Model B extends Model A by including a $\dot H$ term, and can be written as
\begin{equation}
g(t)=\beta_1 H^2+\beta_2 \dot H.
\label{eq:gB_model}
\end{equation}
This form appears in the original entropic-force model and in later related work~\cite{Easson, EFDE8}. 
Compared with Model A, the entropic term depends on both the Hubble parameter and its time variation through the $H^2$ and $\dot H$ terms.

Using
\begin{equation}
\frac{\dot H}{H^2}=-(1+q),
\label{eq:Hdot_q_B}
\end{equation}
Eq.~(\ref{eq:gB_model}) becomes
\begin{equation}
\frac{g(t)}{H^2}=\beta_1-\beta_2(1+q).
\label{eq:g_over_H2_B}
\end{equation}
Substituting Eq.~(\ref{eq:g_over_H2_B}) into Eq.~(\ref{eq:q_general}), we obtain the deceleration parameter
\begin{equation}
q(z)=\frac{\Omega_m(z)+2\Omega_r(z)-2(\beta_1-\beta_2)}{2(1-\beta_2)}.
\label{eq:qB}
\end{equation}
This expression reduces to the Model A result when $\beta_2=0$.

Substituting Eq.~(\ref{eq:g_over_H2_B}) into Eq.~(\ref{eq:wDE_general}), and using Eqs.~(\ref{eq:qB}) and (\ref{eq:closure}), 
we obtain the effective equation of state of the entropic dark energy,
\begin{equation}
w_{\rm DE}(z)=
\frac{\Omega_m(z)+\left(1+\beta_2\right)\Omega_r(z)-1-2\beta_1+3\beta_2}
{3(1-\beta_2)\left[1-\Omega_m(z)-\Omega_r(z)\right]}.
\label{eq:wDEB}
\end{equation}

To derive $E^2(z)$, we combine Eq.~(\ref{eq:qB}) with Eq.~(\ref{eq:qE_relation_A}) and use Eq.~(\ref{eq:OmegaA_sub}). 
This gives a first-order linear differential equation for $E^2(z)$. Imposing the boundary condition $E^2(0)=1$, we obtain
\begin{equation}
\begin{aligned}
E^2(z)={}&\frac{\Omega_{m0}}{1+2\beta_1-3\beta_2}(1+z)^3
+\frac{\Omega_{r0}}{1+\beta_1-2\beta_2}(1+z)^4 \\[3pt]
&+\left(1-\frac{\Omega_{m0}}{1+2\beta_1-3\beta_2}
-\frac{\Omega_{r0}}{1+\beta_1-2\beta_2}\right) \\[3pt]
&\quad\times (1+z)^{\frac{2(1-\beta_1)}{1-\beta_2}} .
\end{aligned}
\label{eq:E2B}
\end{equation}
In Model B, the parameter $\beta_2$ modifies the coefficients of the matter and radiation terms 
and also changes the exponent of the additional contribution. 
This solution is valid for $\beta_2\neq 1$, $1+2\beta_1-3\beta_2\neq 0$, and $1+\beta_1-2\beta_2\neq 0$.

\subsection{Model C}

Model C includes an additional $H^4$ term, and is taken as
\begin{equation}
g(t)=\beta_1 H^2+\beta_2 \dot H+\beta_3 H^4.
\label{eq:gC_model}
\end{equation}
The $H^4$ term can be motivated by logarithmic corrections to the Bekenstein-Hawking
entropy considered in earlier studies.
A contribution proportional to the logarithm of the horizon area was found within the
quantum geometry formulation of the event horizon, while leading-order corrections
arising from small statistical fluctuations around equilibrium were shown to have
a logarithmic dependence on the area
\cite{KaulMajumdar2000,DasMajumdarBhaduri2002}.
In the entropic inflation scenario, including such a logarithmic correction in the
horizon entropy gives rise to an additional $H^4$ term in the acceleration equation,
with a coefficient proportional to $H_{\rm Pl}^{-2}$
\cite{EFDE7}.
In this formulation, the correction relative to the leading $H^2$ term
is proportional to $H^2/H_{\rm Pl}^2$. This scaling makes the correction small at late times
when $H$ is small, while it can become significant in the very early Universe when $H$ is large. 
For the analysis presented here, we retain this previously proposed $H^4$ dependence 
and treat its coefficient $\beta_3$ as a free parameter, so that its possible 
contribution to the late-time expansion can be constrained by the observational data.

Using Eq.~(\ref{eq:Hdot_q_B}) and defining $\tilde{\beta}_3=\beta_3 H_0^2$, Eq.~(\ref{eq:gC_model}) becomes
\begin{equation}
\frac{g(t)}{H^2}=\beta_1-\beta_2(1+q)+\tilde{\beta}_3 E^2(z).
\label{eq:g_over_H2_C}
\end{equation}
Substituting Eq.~(\ref{eq:g_over_H2_C}) into Eq.~(\ref{eq:q_general}), we obtain the deceleration parameter
\begin{equation}
q(z)=\frac{\Omega_m(z)+2\Omega_r(z)-2(\beta_1-\beta_2)-2\tilde{\beta}_3 E^2(z)}{2(1-\beta_2)}.
\label{eq:q_modelC}
\end{equation}
In Model C, the $H^4$ term makes $q(z)$ depend on $E^2(z)$.

For Model C, substituting Eq.~(\ref{eq:g_over_H2_C}) into Eq.~(\ref{eq:wDE_general}), 
and combining the result with Eqs.~(\ref{eq:q_modelC}) and (\ref{eq:closure}), we find
\begin{equation}
\begin{aligned}
w_{\rm DE}(z)
&=\frac{1}{3(1-\beta_2)\left[1-\Omega_m(z)-\Omega_r(z)\right]} \\
&\quad\times \Bigl[\Omega_m(z)+(1+\beta_2)\Omega_r(z)-1-2\beta_1 \\
&\qquad\quad +3\beta_2-2\tilde{\beta}_3 E^2(z)\Bigr].
\end{aligned}
\label{eq:wDE_modelC}
\end{equation}

To derive $E^2(z)$, we combine Eq.~(\ref{eq:q_modelC}) with Eq.~(\ref{eq:qE_relation_A}) and use Eq.~(\ref{eq:OmegaA_sub}). This gives
\begin{equation}
\begin{aligned}
&(1-\beta_2)(1+z)\frac{dE^2}{dz}-2(1-\beta_1)E^2 \\
&\quad+2\tilde{\beta}_3\left(E^2\right)^2
=\Omega_{m0}(1+z)^3+2\Omega_{r0}(1+z)^4.
\end{aligned}
\label{eq:Ez_modelC}
\end{equation}
Because of the quadratic term $\left(E^2\right)^2$, Eq.~(\ref{eq:Ez_modelC}) is a non-homogeneous Riccati equation. 
The evolution of $E^2(z)$ can be obtained numerically.

When the analysis is extended to include CMB distance priors, 
the evolution equation for Model C must be evaluated over a much wider redshift range. 
In that case, solving the exact Riccati equation directly in the parameter fit can lead to additional numerical instability. 
We adopt a first-order perturbative treatment only for this full data combination. 
In the perturbative construction, we set $\Omega_{r0}\approx0$, 
consistently with the effective framework used in the corresponding observational analysis.

Taking the Model B solution with $\Omega_{r0}\approx0$ as the unperturbed solution, 
\begin{equation}
\begin{aligned}
E_B^2(z)
&=\frac{\Omega_{m0}}{1+2\beta_1-3\beta_2}(1+z)^3 \\
&\quad+\left(1-\frac{\Omega_{m0}}{1+2\beta_1-3\beta_2}\right)(1+z)^\delta,
\end{aligned}
\label{eq:EB_modelC}
\end{equation}
with
\begin{equation}
\delta=\frac{2(1-\beta_1)}{1-\beta_2},
\label{eq:Omegaeff_delta_C}
\end{equation}
we write the Model C solution as
\begin{equation}
E_C^2(z)\simeq E_B^2(z)+\tilde{\beta}_3\Delta(z).
\label{eq:Ez_perturbation}
\end{equation}
Here $\Delta(z)$ denotes the leading correction induced by the $H^4$ term. 
Substituting Eq.~(\ref{eq:Ez_perturbation}) into Eq.~(\ref{eq:Ez_modelC}) and retaining only terms linear in $\tilde{\beta}_3$, 
consistent with the perturbative ansatz, we obtain
\begin{equation}
\frac{d\Delta(z)}{dz}-\frac{\delta}{1+z}\Delta(z)
=-\frac{2}{1-\beta_2}\frac{\left[E_B^2(z)\right]^2}{1+z}.
\label{eq:Delta_ODE}
\end{equation}
With the normalization condition $E_C^2(0)=1$, the perturbation satisfies
\begin{equation}
\Delta(0)=0.
\label{eq:Delta_boundary}
\end{equation}

Solving Eq.~(\ref{eq:Delta_ODE}) by the integrating factor method, we obtain
\begin{equation}
\begin{aligned}
\Delta(z)=&\,K(1+z)^\delta \int_0^z \Bigg[
\left(\frac{\Omega_{m0}}{1+2\beta_1-3\beta_2}\right)^2 (1+z')^{5-\delta} \\
&\qquad
+\frac{2\Omega_{m0}}{1+2\beta_1-3\beta_2}
\left(1-\frac{\Omega_{m0}}{1+2\beta_1-3\beta_2}\right) \\
&\qquad\qquad\times (1+z')^2 \\
&\qquad
+\left(1-\frac{\Omega_{m0}}{1+2\beta_1-3\beta_2}\right)^2 \\
&\qquad\qquad\times (1+z')^{\delta-1}
\Bigg]\,dz'.
\end{aligned}
\label{eq:Delta_solution}
\end{equation}
where
\begin{equation}
K=-\frac{2}{1-\beta_2}.
\label{eq:K_C}
\end{equation}
This integral can be evaluated analytically. In particular, for $\delta\neq0$ and $\delta\neq6$, we obtain
\begin{equation}
\begin{aligned}
\Delta(z)=K\Bigg[
&\frac{1}{6-\delta}
\left(\frac{\Omega_{m0}}{1+2\beta_1-3\beta_2}\right)^2(1+z)^6 \\
&+\frac{2}{3}\frac{\Omega_{m0}}{1+2\beta_1-3\beta_2}
\left(1-\frac{\Omega_{m0}}{1+2\beta_1-3\beta_2}\right) \\
&\qquad\times (1+z)^{3+\delta} \\
&+\frac{1}{\delta}
\left(1-\frac{\Omega_{m0}}{1+2\beta_1-3\beta_2}\right)^2(1+z)^{2\delta} \\
&-C_{\Delta}(1+z)^\delta
\Bigg].
\end{aligned}
\end{equation}
with
\begin{equation}
\begin{aligned}
C_{\Delta}=
&\frac{1}{6-\delta}\left(\frac{\Omega_{m0}}{1+2\beta_1-3\beta_2}\right)^2 \\
&+\frac{2}{3}\frac{\Omega_{m0}}{1+2\beta_1-3\beta_2}
\left(1-\frac{\Omega_{m0}}{1+2\beta_1-3\beta_2}\right) \\
&+\frac{1}{\delta}\left(1-\frac{\Omega_{m0}}{1+2\beta_1-3\beta_2}\right)^2.
\end{aligned}
\label{eq:Delta_constant}
\end{equation}
The perturbative treatment is expected to be valid provided that
\begin{equation}
\left|\tilde{\beta}_3\Delta(z)\right|\ll E_B^2(z).
\label{eq:perturbative_validity}
\end{equation}

This perturbative construction is used only in the full data combination considered later. 
It is introduced to capture the main contribution of the $H^4$ term in that case, 
while the exact evolution is retained in the analysis without CMB distance priors. 
In this sense, it should be regarded as an effective treatment for the corresponding fit 
rather than as a complete treatment of the high-redshift regime. 
Its validity is assessed through the condition in Eq.~(\ref{eq:perturbative_validity}).

\subsection{Correspondence Between Theory and Observation}
\label{subsec:correspondence}

In this subsection, we clarify how the matter density parameter is defined 
in the theoretical model and in the observational analysis. 
In the observational fits considered in this work, 
the parameter constrained more directly by the data is the coefficient of the $(1+z)^3$ term in $E^2(z)$, 
rather than the parameter $\Omega_{m0}$ appearing in the theoretical formulation. 
We introduce an effective matter density parameter, which we denote by $\Omega_{\rm eff}$, 
for the observational analysis \cite{EFDE12, effective-matter1}.

For Model A, Eq.~(\ref{eq:E2A}) shows that the coefficient of the $(1+z)^3$ term is
\begin{equation}
\Omega_{\rm eff}= \frac{\Omega_{m0}}{1+2\beta_1}.
\label{eq:Omegaeff_A}
\end{equation}
For Model B, the corresponding coefficient in Eq.~(\ref{eq:E2B}) is
\begin{equation}
\Omega_{\rm eff}= \frac{\Omega_{m0}}{1+2\beta_1-3\beta_2}.
\label{eq:Omegaeff_BC}
\end{equation}
The same combination is used for Model C in the observational analysis. 
This is consistent with the first-order perturbative treatment adopted for the full data combination, 
where the zeroth-order solution is of the Model B form. In the exact treatment, 
the evolution is determined by the Riccati equation in Eq.~(\ref{eq:Ez_modelC}), 
and the full solution is not written in the same explicit form as in Models A and B. 
Accordingly, Eq.~(\ref{eq:Omegaeff_BC}) is the definition adopted in the observational analysis of Model C.

In the observational analysis, 
we use $\Omega_{\rm eff}$ as the matter parameter quoted in the fits. 
The corresponding parameter $\Omega_{m0}$ can be obtained from the above relations once the entropic parameters are determined. 
In this way, $\Omega_{\rm eff}$ is treated as an effective parameter for the cosmological fit, 
rather than as an additional independent matter component.

\section{Theoretical Viability Conditions}
\label{sec:theoretical}

Before comparing the entropic models with observations, 
we examine whether their background evolution satisfies several basic theoretical conditions. 
We consider three aspects. The first is early-universe consistency, 
which requires the entropic dark energy contribution to remain subdominant 
at sufficiently high redshift in order to preserve the standard thermal history. 
The second is the TCC, which in the present cosmological context constrains the future behavior of the models 
by disfavoring eternal accelerated expansion. 
The third is the realization of Quintom-B crossing, motivated by recent DESI-related analyses. 
In the theoretical evolution equations, the matter sector is described by $\Omega_{m0}$, 
whereas the parameter used in the observational fits is $\Omega_{\rm eff}$. 
This distinction is relevant here because the entropic correction terms modify the relation 
between the matter sector in the theoretical description and the expansion history constrained by the data.

\subsection{Early Universe Consistency}
\label{subsec:EUC}

The early-universe consistency condition is introduced to ensure that the entropic driving term does not spoil the standard thermal history of the universe. 
In practice, this means that any effective entropic contribution should remain subdominant at sufficiently high redshift, 
so that the standard radiation-dominated expansion is preserved. 
For the present class of models, 
this requirement can be examined through the asymptotic behavior of the quantity $g(t)/H^2$ in the radiation-dominated limit, 
where $q\rightarrow 1$.

For Model A, using $g(t)/H^2=\beta_1$. 
The early-universe consistency condition requires $|\beta_1|\ll 1$, 
since the entropic contribution must remain negligible in the radiation-dominated regime. 
The same conclusion also follows from Eq.(\ref{eq:qA}): at sufficiently high redshift, 
$\Omega_r\to1$ and $\Omega_m\to0$, so that $q\to1-\beta_1$, 
whereas the standard radiation-dominated limit requires $q\to1$. 
By contrast, late-time acceleration in this model generally favors values of $\beta_1$ that are not much smaller than unity. 
This suggests a tension between early-universe consistency and late-time acceleration in Model A.

For Model B, Eq.(\ref{eq:g_over_H2_B}) implies that in the radiation-dominated limit, 
where $q\to1$, the entropic contribution approaches $g(t)/H^2\to \beta_1-2\beta_2$. 
The early-universe consistency condition requires $|\beta_1-2\beta_2|\ll1$. 
Compared with Model A, this does not force the entropic contribution to remain small at later times, 
since $q$ departs from unity once the universe evolves away from radiation domination. 
In this way, the $\dot H$ term provides additional freedom that can ease the tension with the early-universe consistency requirement.

For Model C, Eq.(\ref{eq:g_over_H2_C}) implies that in the radiation-dominated limit, 
where $q\to1$, the entropic contribution approaches $g(t)/H^2\to \beta_1-2\beta_2+\tilde{\beta}_3E^2(z)$. 
The early-universe consistency condition requires this term to remain subdominant at high redshift. 
Since $E^2(z)$ grows rapidly in the radiation-dominated regime, 
the $H^4$ term can become important even when $\tilde{\beta}_3$ is not large. 
Early-universe consistency constrains $\tilde{\beta}_3$ to remain small.

\subsection{Trans-Planckian Censorship Conjecture in Entropic Cosmology}
\label{subsec:TCC}

The TCC requires that, 
once the effective field theory description is applicable, 
sub-Planckian modes should not be stretched beyond the Hubble radius and classicalize \cite{TCC1, TCC2, TCC3}. 
A convenient way to express this condition is
\begin{equation}
\frac{a(t_f)}{a(t_i)}\,\ell_{\rm Pl}<H^{-1}(t_f).
\label{eq:TCC}
\end{equation}
For the late-time evolution studied here, 
we adopt the corresponding condition that the universe should not remain eternally accelerating \cite{TCC4}. 
For a dark-energy-dominated future, this implies
\begin{equation}
w_{\rm DE}(z\to -1)\ge -\frac{1}{3},
\label{eq:TCC_wDE}
\end{equation}
or equivalently,
\begin{equation}
q_\infty \ge 0.
\label{eq:TCC_q}
\end{equation}

For Model A, Eq.~(\ref{eq:qA}) gives
\begin{equation}
q_\infty=-\beta_1
\label{eq:TCC_ModelA_q}
\end{equation}
in the future limit, where $\Omega_m\to0$ and $\Omega_r\to0$. The TCC condition requires
\begin{equation}
\beta_1\le 0.
\label{eq:TCC_ModelA}
\end{equation}
On the other hand, a late-time accelerating phase requires $q_\infty<0$, which in this model implies $\beta_1>0$. 
This indicates a tension between the TCC condition and late-time acceleration in Model A.

For Model B, Eq.~(\ref{eq:qB}) leads to
\begin{equation}
q_\infty=\frac{\beta_2-\beta_1}{1-\beta_2}
\label{eq:TCC_ModelB_q}
\end{equation}
when $\Omega_m\to0$ and $\Omega_r\to0$. If $1-\beta_2>0$, this gives the TCC condition
\begin{equation}
\beta_2\ge \beta_1.
\label{eq:TCC_ModelB}
\end{equation}
This suggests that the $\dot H$ correction introduces additional freedom for the model to exit the accelerating phase in the far future.

For Model C, Eq.~(\ref{eq:g_over_H2_C}) shows that the $H^4$ term becomes irrelevant in the future for solutions with $H\to0$. 
Under this condition, the TCC constraint takes the same form as in Model B, Eq.~(\ref{eq:TCC_ModelB}). 
This means that, as far as the TCC condition is concerned, 
the late-time viability of Model C is still determined mainly by the lower-order entropic parameters.

These conditions should be understood as asymptotic necessary conditions. 
The full integral form of the TCC may impose a stronger restriction if the transition to $w_{\rm DE}\ge -1/3$ occurs too late. 
Even so, this analysis remains a useful preliminary criterion for the entropic models.

\subsection{Theoretical Realization of Quintom-B Crossing}
\label{subsec:Quintom}

Recent DESI analyses have indicated that the dark energy equation of state 
may evolve from the phantom regime ($w<-1$) to the quintessence regime ($w>-1$) \cite{DESI3, DESI5}. 
This possibility is usually described as Quintom-B crossing \cite{quintom-B1, quintom-B2}. 
In the present framework, this question can be studied through the evolution of $w_{\rm DE}(z)$.

From Eq.~(\ref{eq:wDE_general}) together with Eq.~(\ref{eq:closure}), we can rewrite
\begin{equation}
w_{\rm DE}(z)+1=\frac{2}{3\Omega_{\rm DE}(z)}
\left[\Omega_{\rm DE}(z)-\frac{g(z)}{H^2(z)}\right].
\label{eq:wplus1_general}
\end{equation}
The phantom divide is crossed when
\begin{equation}
\Omega_{\rm DE}(z)=\frac{g(z)}{H^2(z)}.
\label{eq:crossing_condition}
\end{equation}
A Quintom-B transition further requires that $\Omega_{\rm DE}(z)-\frac{g(z)}{H^2(z)}$ 
changes sign from negative to positive as time increases, or equivalently as redshift decreases.

For Model A, the crossing condition becomes
\begin{equation}
\Omega_{\rm DE}(z)=\beta_1.
\label{eq:QuintomB_ModelA_crossing}
\end{equation}
If $\Omega_{\rm DE}(z)$ increases monotonically with time, Model A can exhibit a Quintom-B transition only when
\begin{equation}
0<\beta_1<1.
\label{eq:QuintomB_ModelA_condition}
\end{equation}
Then, the crossing is determined by a fixed value of $\beta_1$, 
so the phantom divide can be crossed at most once. 
If the early-universe consistency condition is also imposed, 
the allowed range is further restricted to small positive values of $\beta_1$. 
The TCC condition requires $\beta_1\le 0$, so the Quintom-B region of Model A is not compatible with the TCC requirement.

For Model B, the crossing condition is
\begin{equation}
\Omega_{\rm DE}(z)=\beta_1-\beta_2\left[1+q(z)\right].
\label{eq:QuintomB_ModelB_crossing}
\end{equation}
In the radiation-dominated limit, where $q\to 1$ and $\Omega_{\rm DE}\to 0$, a Quintom-B history requires
\begin{equation}
\beta_1-2\beta_2>0,
\label{eq:QuintomB_ModelB_early}
\end{equation}
so that the evolution starts on the phantom side. 
In the future limit, where $\Omega_{\rm DE}\to 1$, the late-time quintessence side requires
\begin{equation}
\frac{1-\beta_1}{1-\beta_2}>0.
\label{eq:QuintomB_ModelB_future}
\end{equation}
If $1-\beta_2>0$, this reduces to $\beta_1<1$. Compared with Model A, 
the inclusion of the $\dot H$ term makes the crossing condition time dependent 
and gives the model more freedom to realize a Quintom-B transition. 
If the early-universe consistency condition is also imposed, the relation above is further restricted to
\begin{equation}
0<\beta_1-2\beta_2\ll 1.
\label{eq:QuintomB_ModelB_overlap}
\end{equation}
The TCC condition restricts the remaining overlap region, but it is not as restrictive as in Model A.

For Model C, the corresponding crossing condition is
\begin{equation}
\Omega_{\rm DE}(z)=\beta_1-\beta_2\left[1+q(z)\right]+\tilde{\beta}_3E^2(z).
\label{eq:QuintomB_ModelC_crossing}
\end{equation}
At early times, when $\Omega_{\rm DE}\to 0$ and $q\to 1$, Quintom-B history requires
\begin{equation}
\beta_1-2\beta_2+\tilde{\beta}_3E^2(z)>0.
\label{eq:QuintomB_ModelC_early}
\end{equation} 
In the far future, for solutions with $H\to 0$, the $H^4$ term becomes negligible, 
and the late-time condition again reduces to the Model B form, Eq.(\ref{eq:QuintomB_ModelB_future}). 
Thus, Model C can also exhibit a Quintom-B transition, but the crossing is shifted by the extra $H^4$ contribution. 
Once the early-universe consistency condition is also imposed, $\tilde{\beta}_3$ is constrained to remain small, 
so the viable Quintom-B region of Model C is still controlled mainly by the lower-order parameters,  
with the $H^4$ term providing a correction. 
The TCC condition further reduces the allowed overlap region, 
but does not remove the possibility of Quintom-B behavior.

\section{Observational Constraints}
\label{sec:observations}

To constrain the entropic models, we perform a joint likelihood analysis 
and estimate the model parameters through Markov Chain Monte Carlo (MCMC) sampling \cite{MCMC1, MCMC2}, 
implemented with the emcee package \cite{emcee}. 
As discussed in Sec.~\ref{sec:models}, the effective matter density $\Omega_{\rm eff}$ is taken as the primary sampling parameter, 
since it is more directly connected with the observable expansion history. 
We neglect the radiation contribution, $\Omega_{r0}\approx0$, 
to keep the observational analysis consistent with the model derivation in Sec.~\ref{sec:models}. 
The total chi-squared is written as
\begin{equation}
\chi^2_{\rm tot}=\chi^2_{\rm SN}+\chi^2_{\rm BAO}+\chi^2_{\rm CMB}.
\end{equation}

\subsection{Type Ia Supernovae: Pantheon+}

We use the Pantheon+ SN Ia sample \cite{pantheonPLUS1,pantheonPLUS2}, 
which contains 1701 light curves from 1550 Type Ia supernovae over the redshift range $0.001 \le z \le 2.26$. 
The theoretical apparent magnitude is written as
\begin{equation}
m_{B,\rm th}(z)=5\log_{10}\left[\frac{d_L(z)}{10\,\mathrm{pc}}\right]+M_B,
\end{equation}
where $d_L(z)$ is the luminosity distance and $M_B$ is the absolute magnitude. 
To reduce the dependence on the absolute magnitude calibration, we analytically marginalize over $M_B$ \cite{Mb}. 
The corresponding chi-squared is
\begin{equation}
\chi^2_{\rm SN}=\mathcal{S}_2-\frac{\mathcal{S}_1^2}{\mathcal{S}_0}+\ln\left(\frac{\mathcal{S}_0}{2\pi}\right),
\end{equation}
with
\begin{align}
\mathcal{S}_0 &= \sum_{i,j} C^{-1}_{{\rm SN},ij}, \\
\mathcal{S}_1 &= \sum_{i,j} C^{-1}_{{\rm SN},ij}
\left[m_{B,\mathrm{obs},i}-5\log_{10}\left(\frac{d_L(z_i)}{10\,\mathrm{pc}}\right)\right], \\
\mathcal{S}_2 &= \sum_{i,j}
\left[m_{B,\mathrm{obs},i}-5\log_{10}\left(\frac{d_L(z_i)}{10\,\mathrm{pc}}\right)\right]
\nonumber\\
&\qquad C^{-1}_{{\rm SN},ij}
\left[m_{B,\mathrm{obs},j}-5\log_{10}\left(\frac{d_L(z_j)}{10\,\mathrm{pc}}\right)\right].
\end{align}
Here $C_{\rm SN}$ denotes the full covariance matrix, including both statistical and systematic uncertainties.

\subsection{Baryon Acoustic Oscillations: DESI DR2}

DESI DR2 provides BAO measurements over the redshift range $0.1 < z < 4.2$, 
using galaxies, quasars, and the Lyman-$\alpha$ forest \cite{DESI2,DESI3}. 
These measurements constrain cosmic distances through the sound horizon at the drag epoch,
\begin{equation}
r_d=\int_{z_d}^{\infty}\frac{c_s(z)}{H(z)}\,dz,
\end{equation}
where $c_s(z)$ is the photon-baryon sound speed and $z_d$ is the drag redshift. 
The transverse comoving distance is given by
\begin{equation}
D_M(z)=c\int_0^z\frac{dz'}{H(z')}.
\end{equation}
The Hubble distance is
\begin{equation}
D_H(z)=\frac{c}{H(z)}.
\end{equation}
For isotropic BAO measurements, we use the volume-averaged distance
\begin{equation}
D_V(z)=\left[zD_H(z)D_M^2(z)\right]^{1/3}.
\end{equation}
The directly constrained observables are $D_M(z)/r_d$, $D_H(z)/r_d$, and $D_V(z)/r_d$. 
In our analysis, $r_d$ is treated as an external standard ruler. 
The corresponding chi-squared is constructed from the DESI DR2 measurements and covariance matrix.

\subsection{CMB Distance Priors}

To complement the SN Ia and BAO data, we use the CMB distance priors from Planck 2018 \cite{CMB}. 
This compressed description replaces the full CMB likelihood with four parameters and their associated $4\times4$ covariance matrix, 
while retaining the main constraining power on the distance scale and matter content. 
It provides an efficient way to include CMB information in the analysis with lower computational cost.

The observational vector is written as
\begin{equation}
\boldsymbol{V}_{\rm CMB}=
\begin{pmatrix}
R \\[4pt]
\ell_a \\[4pt]
\Omega_b h^2 \\[4pt]
n_s
\end{pmatrix},
\end{equation}
and the corresponding theoretical prediction is denoted by $\boldsymbol{V}_{\rm CMB}^{\rm th}$, 
where $R$ is the shift parameter, $\ell_a$ is the acoustic angular scale, 
$\Omega_b h^2$ is the physical baryon density, and $n_s$ is the scalar spectral index.

The CMB chi-squared is given by
\begin{equation}
\chi^2_{\rm CMB}
=
(\boldsymbol{V}_{\rm CMB}-\boldsymbol{V}_{\rm CMB}^{\rm th})^{\rm T}
\mathbf{C}_{\rm CMB}^{-1}
(\boldsymbol{V}_{\rm CMB}-\boldsymbol{V}_{\rm CMB}^{\rm th}),
\end{equation}
where $\mathbf{C}_{\rm CMB}$ is the covariance matrix of the compressed CMB data.

\subsection{Statistical Inference and Model Comparison}

We sample the posterior distributions using the emcee package. 
To assess the relative statistical performance of the models, we use the AIC, defined as
\begin{equation}
\mathrm{AIC}=\chi^2_{\rm min}+2k,
\end{equation}
where $k$ is the number of free parameters.  
We compute $\Delta\mathrm{AIC}$ values with Model A as the reference model.

\subsection{Observational Results}

\begin{figure*}[ht!]\suppressfloats
\centering
\includegraphics [width=5.5cm]{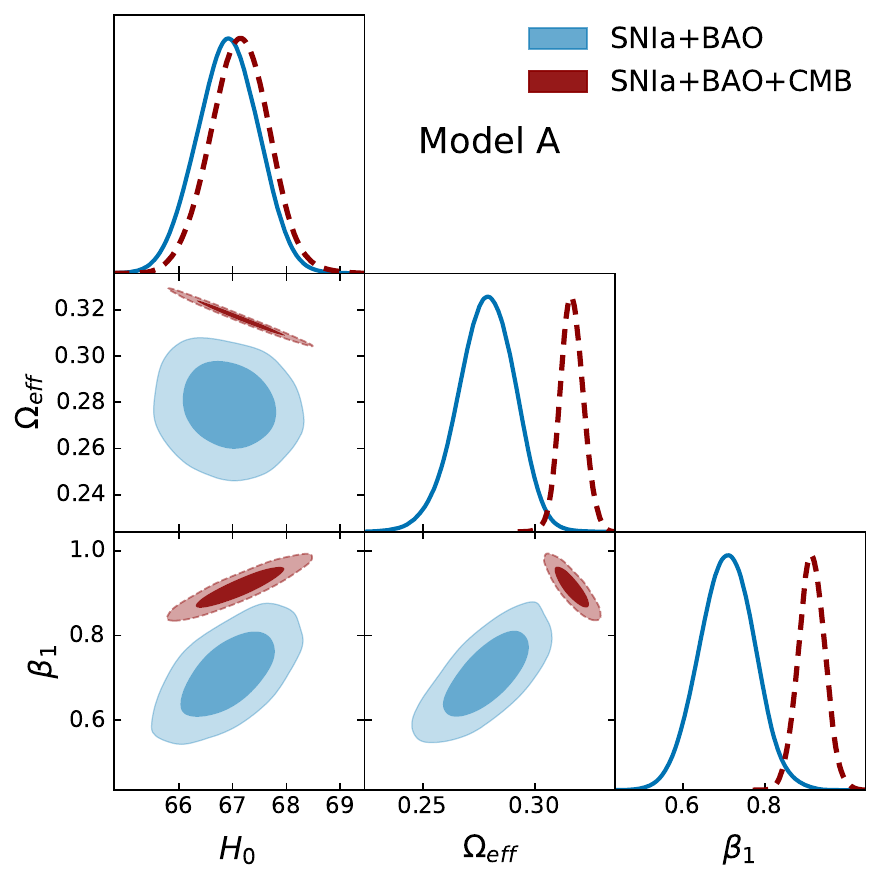}
\includegraphics [width=5.5cm]{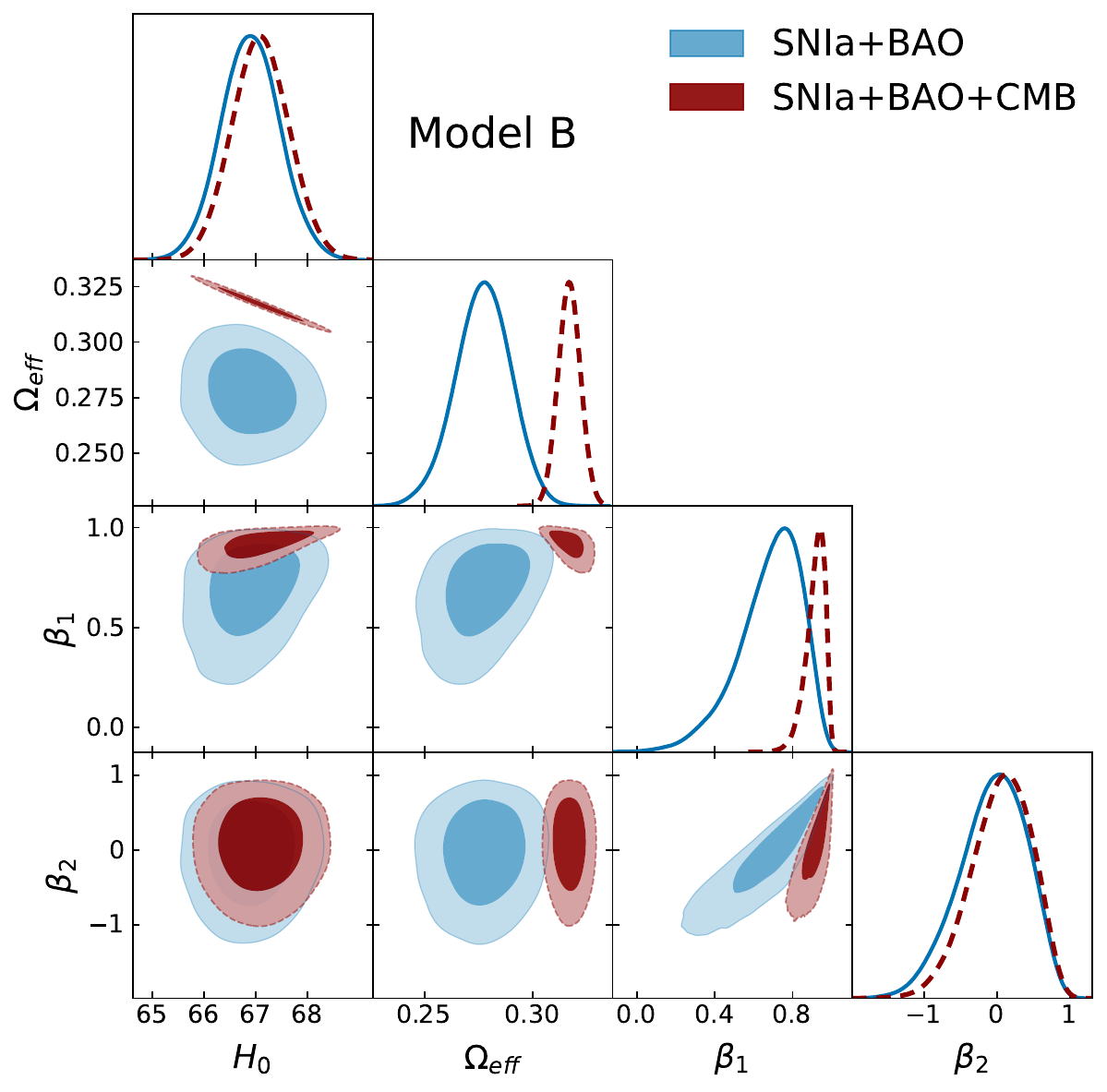}
\includegraphics [width=5.5cm]{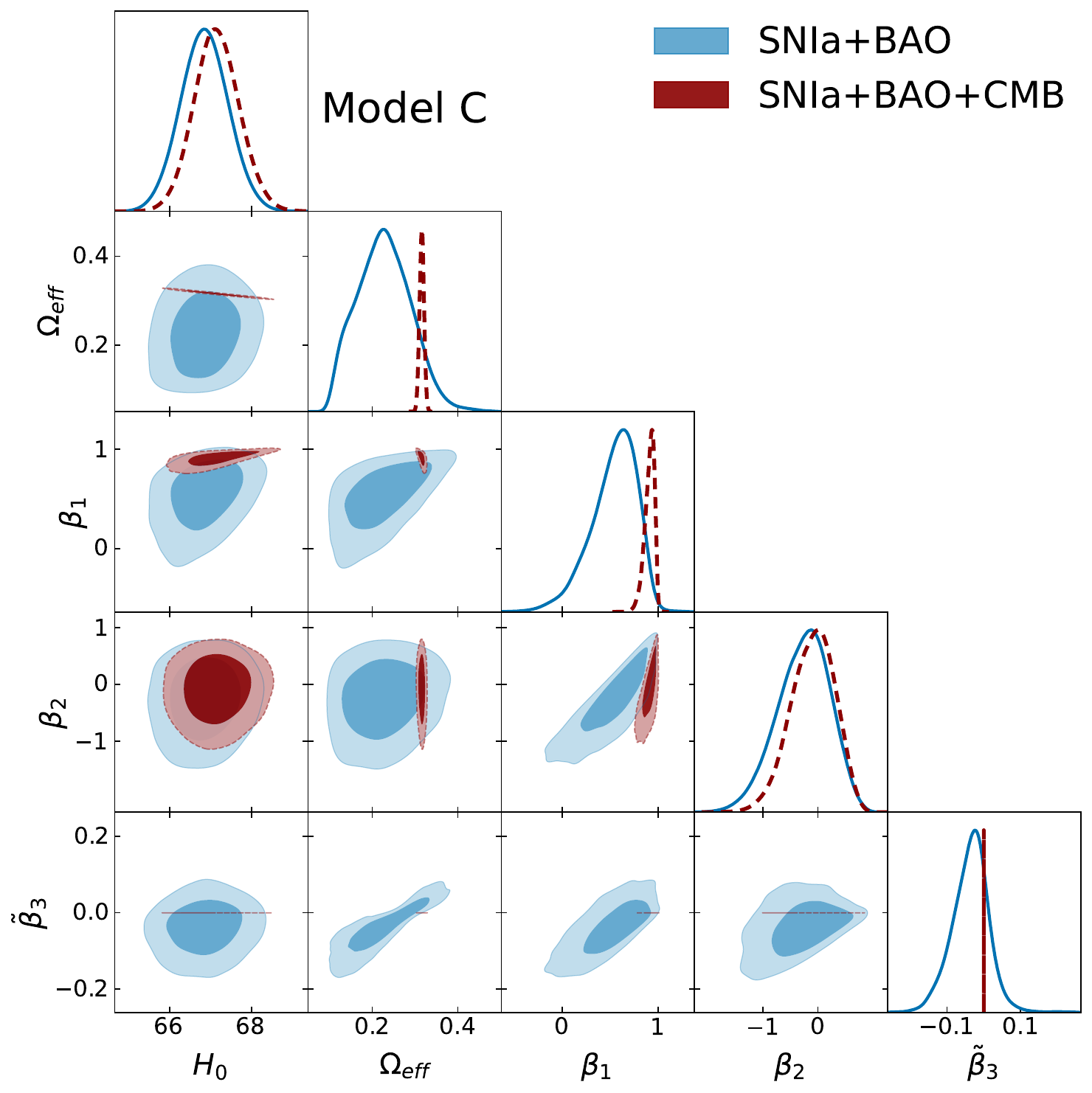}   
\caption{Marginalized posterior distributions and contours at the 68\% and 95\% confidence levels for Models A, B, and C. 
The light blue curves and contours correspond to the SNIa+BAO constraints, while the dark red curves and contours correspond to the joint SNIa+BAO+CMB constraints.}
\label{fig-getdist}
\end{figure*}

\begin{table*}
\centering   
\resizebox{\textwidth}{!}{  
\begin{tabular}{l c c c c c c c c} 
 \hline
 Model & $H_0$  &  $\Omega_{eff}$ &  $\Omega_b$ & $\beta_1$ & $\beta_2$ & $\tilde{\beta}_3$ & $\chi^2_\mathrm{tot}$ & $\Delta \mathrm{AIC}$ \\
 \hline
$\text{Model A}_{\text{SNIa+BAO}}$ & $ 66.9129 \pm 0.5755 $ & $ 0.2786 \pm 0.0124 $ & - & $ 0.7078 \pm 0.0661 $ & - & - &  1765.91 & 0 \\
$\text{Model B}_{\text{SNIa+BAO}}$ & $ 66.8993 \pm 0.5714 $ & $ 0.2773 \pm 0.0125 $ & - & $ 0.7079^{+0.1368}_{-0.1868}$ & $ -0.0206^{+0.4460}_{-0.5079} $ & -  &  1765.92  &  2.01  \\
$\text{Model C}_{\text{SNIa+BAO}}$ & $ 66.8523 \pm 0.5782 $ & $ 0.2264 \pm 0.0658 $ & - & $ 0.5697^{+0.2043}_{-0.2793}$ & $ -0.2308^{+0.4479}_{-0.5247} $ & $ -0.0334^{+0.0423}_{-0.0525} $ &  1768.98  &  7.07  \\
\hline
$\text{Model A}_{\text{SNIa+BAO+CMB}}$ & $ 67.1341 \pm 0.5502 $ & $ 0.3165 \pm 0.0052 $ & $ 0.0496 \pm 0.0008 $ & $ 0.9133 \pm 0.0325 $ & - & - & 1778.56 & 0  \\
$\text{Model B}_{\text{SNIa+BAO+CMB}}$ & $ 67.0916 \pm 0.5559 $ & $ 0.3170 \pm 0.0052 $ & $ 0.0497 \pm 0.0009 $ & $ 0.9251^{+0.0401}_{-0.0584}$ & $ 0.0852^{+0.4014 }_{-0.4498} $ & - &  1778.62  & 2.06 \\
$\text{Model C}_{\text{SNIa+BAO+CMB}}$ & $ 67.1405 \pm 0.5602 $ & $ 0.3160 \pm 0.0051 $ & $ 0.0497 \pm 0.0009 $ & $ 0.9171^{+0.0423}_{-0.0616}$ & $ -0.0686^{+0.3813}_{-0.4333} $ & $(-20.0590^{+26.5662}_{-36.0682} ) \times 10^{-10}$ & 1778.33 & 3.77 \\
\hline
\end{tabular}
}
\caption{Best-fit parameters and their $1\sigma$ uncertainties for the entropic models shown in Fig.~\ref{fig-getdist}. 
The $\Delta \mathrm{AIC}$ is defined relative to Model A within each data combination, with larger values indicating weaker statistical support. 
For Model C in the SNIa+BAO+CMB combination, $\tilde{\beta}_3$ is reported in units of $10^{-10}$ for readability.}
\label{table-data}
\end{table*}

The best-fit values and their $1\sigma$ uncertainties are summarized in Table\ref{table-data}, 
and the marginalized posterior distributions are shown in Fig.\ref{fig-getdist}. 
When the CMB distance priors are included, the constraints on the main parameters become significantly tighter, 
particularly for $\Omega_{\rm eff}$ and $\beta_1$. 
For the SNIa+BAO combination alone, the parameter degeneracies remain broader.

Including the CMB distance priors shifts the preferred region toward higher $\Omega_{\rm eff}$ and larger $\beta_1$ in all three models. 
In Model A, $\Omega_{\rm eff}$ moves from about $0.28$ to about $0.32$, while $\beta_1$ increases from about $0.71$ to about $0.91$. 
The same tendency appears in Models B and C. 
The value of $H_0$ changes only slightly after the CMB priors are included. 
This suggests that the high-redshift geometric information mainly favors a larger effective matter density and a larger entropic contribution.

For the additional correction terms, the current data do not provide clear evidence for a nonzero contribution. 
In both Model B and Model C, the parameter $\beta_2$ remains consistent with zero within $1\sigma$ for both data combinations. 
For Model C, $\tilde{\beta}_3$ is only weakly constrained by SNIa+BAO alone, 
while the inclusion of the CMB priors leads to a much tighter constraint and still keeps it consistent with zero, 
with a magnitude of order $10^{-9}$. 
This suggests that higher-order entropic corrections are not favored by the present data 
and are constrained to remain small once CMB information is included.
The constraint on $\tilde{\beta}_3$ also allows the possible $H^4$ 
contribution in Model C to be examined using the current observational data.

According to the AIC, Model A remains the statistically preferred model for both data combinations, given its smaller number of parameters. 
Model B gives $\Delta \mathrm{AIC}=2.01$ for SNIa+BAO and $\Delta \mathrm{AIC}=2.06$ for SNIa+BAO+CMB, 
so it is not preferred over Model A, although the difference remains modest. 
Model C is less favored by the SNIa+BAO data, with $\Delta \mathrm{AIC}=7.07$, 
while the inclusion of the CMB priors reduces this value to $\Delta \mathrm{AIC}=3.77$. 
This indicates that the higher-order correction does not lead to a sufficient statistical improvement once the extra parameter is included.

Using the relations between $\Omega_{\rm eff}$ and $\Omega_{m0}$ given in 
Sec.~\ref{subsec:correspondence}, the best-fit parameters yield $q_0<0$ 
for all three models and for both data combinations, corresponding to accelerated expansion 
at the present epoch. 
A comparison with the theoretical conditions discussed in Sec.~\ref{sec:theoretical} shows 
that the parameter regions preferred by the data do not fully overlap with those selected 
by the early-universe consistency and TCC conditions.
Among the three models, Model B has the mildest tension between observational viability and theoretical consistency, 
since it remains statistically competitive while still allowing additional dynamical freedom through the $\dot H$ term.
Model A is more tightly constrained by the theoretical conditions. 
For Model C, the data favor a region in which $\tilde{\beta}_3$ remains close to zero, 
so that the higher-order correction enters only as a small contribution.

\section{Conclusion}
\label{sec:conclusion}

In this work, we investigated entropic cosmology by considering three representative forms of the entropic driving term, 
denoted as Model A, Model B, and Model C, with contributions proportional to $H^2$, $H^2+\dot H$, and $H^2+\dot H+H^4$, respectively. 
Within a unified effective dark energy description, we derived the corresponding expressions for the effective equation of state, 
the deceleration parameter, and the expansion history of each model. 
This framework allows the three models to be examined under both theoretical consistency conditions and observational constraints 
within the same effective description.

The theoretical analysis focused on early universe consistency, the TCC condition, and the realization of Quintom-B crossing. 
The results suggest that Model A is the most restricted model. 
Although it gives the simplest expansion history, it is difficult for it to satisfy both the early universe requirements and the TCC condition at the same time. 
Models B and C allow more flexibility through their additional dynamical terms. 
In Model B, the $\dot H$ term can help reduce the entropic contribution at high redshift and can also support a future transition to a non-accelerating phase. 
In Model C, the higher-order term provides an additional correction, although its allowed range remains limited by the theoretical conditions.

Using Pantheon+, DESI DR2, and the Planck 2018 CMB distance priors, 
we found that Model A remains statistically preferred, with fewer free parameters than the other models.
Model B is not preferred over Model A by the information criteria, but it remains statistically competitive 
and shows the mildest tension between observational viability and theoretical consistency. 
For Model C, the higher-order parameter $\tilde{\beta}_3$ is driven close to zero once the CMB information is included, 
indicating that the $H^4$ term can only provide a small correction.

These results show that entropic cosmology can reproduce viable expansion histories, 
but the regions preferred by theoretical consistency and by observations do not fully overlap. 
Among the three models, Model B is the most favorable, 
since it retains additional dynamical freedom while remaining statistically competitive with current data.

The present conclusions concern the theoretical consistency 
and observational viability of the models within the FLRW description adopted here, 
while the corresponding structure-growth behavior requires a dedicated perturbation analysis. 
Such an extension would provide a more complete assessment of whether the viability indicated by the expansion history 
is also maintained in the evolution of cosmological perturbations.

\section*{Acknowledgements}

This work was supported by the National SKA Program of China 
(Grants Nos. 2022SKA0110200 and 2022SKA0110203) and National Natural Science Foundation of China (No.12463001). 
This work was also supported by Xiaofeng Yang's ZHISHAN Distinguished Professor startup funding of Henan University.

\bibliographystyle{elsarticle-num} 
\bibliography{EFDE-reference}

@article{acceleration1,
doi = {10.1086/300499},
url = {https://doi.org/10.1086/300499},
year = {1998},
month = {sep},
publisher = {},
volume = {116},
number = {3},
pages = {1009},
author = {Riess, Adam G. and Filippenko, Alexei V. and Challis, Peter and Clocchiatti, Alejandro and Diercks, Alan and Garnavich, Peter M. and Gilliland, Ron L. and Hogan, Craig J. and Jha, Saurabh and Kirshner, Robert P. and Leibundgut, B. and Phillips, M. M. and Reiss, David and Schmidt, Brian P. and Schommer, Robert A. and Smith, R. Chris and Spyromilio, J. and Stubbs, Christopher and Suntzeff, Nicholas B. and Tonry, John},
title = {Observational Evidence from Supernovae for an Accelerating Universe and a Cosmological Constant},
journal = {The Astronomical Journal}
}

@ARTICLE{acceleration2,
       author = {{Perlmutter}, S. and {Aldering}, G. and {della Valle}, M. and {Deustua}, S. and {Ellis}, R.~S. and {Fabbro}, S. and {Fruchter}, A. and {Goldhaber}, G. and {Groom}, D.~E. and {Hook}, I.~M. and {Kim}, A.~G. and {Kim}, M.~Y. and {Knop}, R.~A. and {Lidman}, C. and {McMahon}, R.~G. and {Nugent}, P. and {Pain}, R. and {Panagia}, N. and {Pennypacker}, C.~R. and {Ruiz-Lapuente}, P. and {Schaefer}, B. and {Walton}, N.},
        title = "{Discovery of a supernova explosion at half the age of the Universe}",
      journal = {Nature},
         year = 1998,
        month = jan,
       volume = {391},
       number = {6662},
        pages = {51-54},
          doi = {10.1038/34124},
archivePrefix = {arXiv},
       eprint = {astro-ph/9712212},
 primaryClass = {astro-ph},
       adsurl = {https://ui.adsabs.harvard.edu/abs/1998Natur.391...51P}
}

@article{acceleration3,
    author = "Perlmutter, S. and others",
    collaboration = "Supernova Cosmology Project",
    title = "{Measurements of $\Omega$ and $\Lambda$ from 42 High Redshift Supernovae}",
    eprint = "astro-ph/9812133",
    archivePrefix = "arXiv",
    reportNumber = "LBNL-41801, LBL-41801",
    doi = "10.1086/307221",
    journal = "Astrophys. J.",
    volume = "517",
    pages = "565--586",
    year = "1999"
}

@article{acceleration4,
  title = {Constraining Dark Energy with Type Ia Supernovae and Large-Scale Structure},
  author = {Perlmutter, Saul and Turner, Michael S. and White, Martin},
  journal = {Phys. Rev. Lett.},
  volume = {83},
  issue = {4},
  pages = {670--673},
  numpages = {0},
  year = {1999},
  month = {Jul},
  publisher = {American Physical Society},
  doi = {10.1103/PhysRevLett.83.670},
  url = {https://link.aps.org/doi/10.1103/PhysRevLett.83.670}
}

@article{LCDM-problem1,
  title = {The cosmological constant problem},
  author = {Weinberg, Steven},
  journal = {Rev. Mod. Phys.},
  volume = {61},
  issue = {1},
  pages = {1--23},
  numpages = {0},
  year = {1989},
  month = {Jan},
  publisher = {American Physical Society},
  doi = {10.1103/RevModPhys.61.1},
  url = {https://link.aps.org/doi/10.1103/RevModPhys.61.1}
}

@ARTICLE{LCDM-problem2,
       author = {{Carroll}, Sean M.},
        title = "{The Cosmological Constant}",
      journal = {Living Reviews in Relativity},
         year = 2001,
        month = dec,
       volume = {4},
       number = {1},
          eid = {1},
        pages = {1},
          doi = {10.12942/lrr-2001-1},
archivePrefix = {arXiv},
       eprint = {astro-ph/0004075},
 primaryClass = {astro-ph},
       adsurl = {https://ui.adsabs.harvard.edu/abs/2001LRR.....4....1C}
}

@article{dynamical1,
title = {A model of holographic dark energy},
journal = {Physics Letters B},
volume = {603},
number = {1},
pages = {1-5},
year = {2004},
issn = {0370-2693},
doi = {https://doi.org/10.1016/j.physletb.2004.10.014},
author = {Miao Li}
}

@article{dynamical2,
doi = {10.1088/1475-7516/2004/08/013},
url = {https://dx.doi.org/10.1088/1475-7516/2004/08/013},
year = {2004},
month = {aug},
publisher = {},
volume = {2004},
number = {08},
pages = {013},
author = {Qing-Guo Huang and Miao Li},
title = {The holographic dark energy in a non-flat universe},
journal = {Journal of Cosmology and Astroparticle Physics}
}

@article{dynamical3,
title = {Holographic dark energy},
journal = {Physics Reports},
volume = {696},
pages = {1-57},
year = {2017},
note = {Holographic Dark Energy},
issn = {0370-1573},
doi = {https://doi.org/10.1016/j.physrep.2017.06.003},
author = {Shuang Wang and Yi Wang and Miao Li}
}

@article{dynamical4,
title = {A dark energy model characterized by the age of the Universe},
journal = {Physics Letters B},
volume = {657},
number = {4},
pages = {228-231},
year = {2007},
issn = {0370-2693},
doi = {https://doi.org/10.1016/j.physletb.2007.09.061},
author = {Rong-Gen Cai}
}

@article{dynamical5,
title = {A new model of agegraphic dark energy},
journal = {Physics Letters B},
volume = {660},
number = {3},
pages = {113-117},
year = {2008},
issn = {0370-2693},
doi = {https://doi.org/10.1016/j.physletb.2007.12.030},
author = {Hao Wei and Rong-Gen Cai}
}

@article{dynamical6,
  title = {Holographic dark energy model from Ricci scalar curvature},
  author = {Gao, Changjun and Wu, Fengquan and Chen, Xuelei and Shen, You-Gen},
  journal = {Phys. Rev. D},
  volume = {79},
  issue = {4},
  pages = {043511},
  numpages = {7},
  year = {2009},
  month = {Feb},
  publisher = {American Physical Society},
  doi = {10.1103/PhysRevD.79.043511},
  url = {https://link.aps.org/doi/10.1103/PhysRevD.79.043511}
}

@article{dynamical7,
  title = {Tsallis holographic dark energy for inflation},
  author = {Mohammadi, Abolhassan and Golanbari, Tayeb and Bamba, Kazuharu and Lobo, Iarley P.},
  journal = {Phys. Rev. D},
  volume = {103},
  issue = {8},
  pages = {083505},
  numpages = {8},
  year = {2021},
  month = {Apr},
  publisher = {American Physical Society},
  doi = {10.1103/PhysRevD.103.083505},
  url = {https://link.aps.org/doi/10.1103/PhysRevD.103.083505}
}

@article{dynamical8,
    author = "Dheepika, M. and Mathew, Titus K.",
    title = "{Tsallis holographic dark energy reconsidered}",
    eprint = "2110.04031",
    archivePrefix = "arXiv",
    primaryClass = "gr-qc",
    doi = "10.1140/epjc/s10052-022-10365-2",
    journal = "Eur. Phys. J. C",
    volume = "82",
    number = "5",
    pages = "399",
    year = "2022"
}

@article{dynamical9,
title = {Tsallis holographic dark energy},
journal = {Physics Letters B},
volume = {781},
pages = {195-200},
year = {2018},
issn = {0370-2693},
doi = {https://doi.org/10.1016/j.physletb.2018.04.001},
author = {M. Tavayef and A. Sheykhi and Kazuharu Bamba and H. Moradpour}
}

@article{dynamical10,
  title = {Holographic dark energy from nonadditive entropy: Cosmological perturbations and observational constraints},
  author = {D'Agostino, Rocco},
  journal = {Phys. Rev. D},
  volume = {99},
  issue = {10},
  pages = {103524},
  numpages = {10},
  year = {2019},
  month = {May},
  publisher = {American Physical Society},
  doi = {10.1103/PhysRevD.99.103524},
  url = {https://link.aps.org/doi/10.1103/PhysRevD.99.103524}
}

@ARTICLE{dynamical11,
       author = {{Moradpour}, H. and {Moosavi}, S.~A. and {Lobo}, I.~P. and {Gra{\c{c}}a}, J.~P. Morais and {Jawad}, A. and {Salako}, I.~G.},
        title = "{Thermodynamic approach to holographic dark energy and the R{\'e}nyi entropy}",
      journal = {European Physical Journal C},
         year = 2018,
        month = oct,
       volume = {78},
       number = {10},
          eid = {829},
        pages = {829},
          doi = {10.1140/epjc/s10052-018-6309-8},
archivePrefix = {arXiv},
       eprint = {1803.02195},
 primaryClass = {physics.gen-ph},
       adsurl = {https://ui.adsabs.harvard.edu/abs/2018EPJC...78..829M}
}

@article{dynamical12,
  title = {Barrow holographic dark energy},
  author = {Saridakis, Emmanuel N.},
  journal = {Phys. Rev. D},
  volume = {102},
  issue = {12},
  pages = {123525},
  numpages = {5},
  year = {2020},
  month = {Dec},
  publisher = {American Physical Society},
  doi = {10.1103/PhysRevD.102.123525},
  url = {https://link.aps.org/doi/10.1103/PhysRevD.102.123525}
}

@article{dynamical13,
title = {Generalized entropy formalism and a new holographic dark energy model},
journal = {Physics Letters B},
volume = {780},
pages = {21-24},
year = {2018},
issn = {0370-2693},
doi = {https://doi.org/10.1016/j.physletb.2018.02.052},
author = {A. {Sayahian Jahromi} and S.A. Moosavi and H. Moradpour and J.P. {Morais Graça} and I.P. Lobo and I.G. Salako and A. Jawad}
}

@article{Easson,
title = {Entropic accelerating universe},
journal = {Physics Letters B},
volume = {696},
number = {3},
pages = {273-277},
year = {2011},
issn = {0370-2693},
doi = {https://doi.org/10.1016/j.physletb.2010.12.025},
author = {Damien A. Easson and Paul H. Frampton and George F. Smoot}
}

@ARTICLE{Verlinde,
       author = {{Verlinde}, Erik},
        title = "{On the origin of gravity and the laws of Newton}",
      journal = {Journal of High Energy Physics},
         year = 2011,
        month = apr,
       volume = {2011},
          eid = {29},
        pages = {29},
          doi = {10.1007/JHEP04(2011)029},
archivePrefix = {arXiv},
       eprint = {1001.0785},
 primaryClass = {hep-th},
       adsurl = {https://ui.adsabs.harvard.edu/abs/2011JHEP...04..029V}
}

@article{EFDE1,
author = {Naeem, Muhammad and Bibi, A.},
year = {2023},
month = {01},
pages = {442},
title = {Accelerating universe via entropic models},
volume = {138},
journal = {The European Physical Journal Plus},
doi = {10.1140/epjp/s13360-023-04073-3}
}

@article{EFDE2,
  title = {CMB acoustic scale in the entropiclike accelerating universe},
  author = {Casadio, R. and Gruppuso, A.},
  journal = {Phys. Rev. D},
  volume = {84},
  issue = {2},
  pages = {023503},
  numpages = {5},
  year = {2011},
  month = {Jul},
  publisher = {American Physical Society},
  doi = {10.1103/PhysRevD.84.023503},
  url = {https://link.aps.org/doi/10.1103/PhysRevD.84.023503}
}

@article{EFDE3,
doi = {10.1088/1475-7516/2011/02/027},
url = {https://doi.org/10.1088/1475-7516/2011/02/027},
year = {2011},
month = {feb},
publisher = {},
volume = {2011},
number = {02},
pages = {027},
author = {Tomi S. Koivisto and David F. Mota and Miguel Zumalacárregui},
title = {Constraining entropic cosmology},
journal = {Journal of Cosmology and Astroparticle Physics}
}

@article{EFDE4,
author = {Gohar, Hussain and Salzano, Vincenzo},
year = {2021},
month = {04},
pages = {},
title = {Cosmological constraints on entropic cosmology with matter creation},
volume = {81},
journal = {The European Physical Journal C},
doi = {10.1140/epjc/s10052-021-09086-9}
}

@article{EFDE5,
title = {Cosmological constraints on the modified entropic force model},
journal = {Physics Letters B},
volume = {692},
number = {3},
pages = {167-175},
year = {2010},
issn = {0370-2693},
doi = {https://doi.org/10.1016/j.physletb.2010.07.036},
author = {Hao Wei}
}

@article{EFDE6,
  title = {Entropic cosmology for a generalized black-hole entropy},
  author = {Komatsu, Nobuyoshi and Kimura, Shigeo},
  journal = {Phys. Rev. D},
  volume = {88},
  issue = {8},
  pages = {083534},
  numpages = {16},
  year = {2013},
  month = {Oct},
  publisher = {American Physical Society},
  doi = {10.1103/PhysRevD.88.083534},
  url = {https://link.aps.org/doi/10.1103/PhysRevD.88.083534}
}

@article{EFDE7,
author = {EASSON, DAMIEN A. and FRAMPTON, PAUL H. and SMOOT, GEORGE F.},
title = {ENTROPIC INFLATION},
journal = {International Journal of Modern Physics A},
volume = {27},
number = {12},
pages = {1250066},
year = {2012},
doi = {10.1142/S0217751X12500662},

URL = { 
    
        https://doi.org/10.1142/S0217751X12500662
    
    

},
eprint = { 
    
        https://doi.org/10.1142/S0217751X12500662
    
    

}
}

@article{EFDE8,
  title = {Entropic-force dark energy reconsidered},
  author = {Basilakos, Spyros and Sol\`a, Joan},
  journal = {Phys. Rev. D},
  volume = {90},
  issue = {2},
  pages = {023008},
  numpages = {11},
  year = {2014},
  month = {Jul},
  publisher = {American Physical Society},
  doi = {10.1103/PhysRevD.90.023008},
  url = {https://link.aps.org/doi/10.1103/PhysRevD.90.023008}
}

@article{EFDE9,
    author = "Mathew, Titus K. and Murali, Chinthak and Shejeelammal, J.",
    title = "{Evolution of non-interacting entropic dark energy and its phantom nature}",
    eprint = "1506.02123",
    archivePrefix = "arXiv",
    primaryClass = "gr-qc",
    doi = "10.1142/S0217732316500711",
    journal = "Mod. Phys. Lett. A",
    volume = "31",
    number = "12",
    pages = "1650071",
    year = "2016"
}

@article{EFDE10,
  title = {Evolution of the universe in entropic cosmologies via different formulations},
  author = {Komatsu, Nobuyoshi and Kimura, Shigeo},
  journal = {Phys. Rev. D},
  volume = {89},
  issue = {12},
  pages = {123501},
  numpages = {22},
  year = {2014},
  month = {Jun},
  publisher = {American Physical Society},
  doi = {10.1103/PhysRevD.89.123501},
  url = {https://link.aps.org/doi/10.1103/PhysRevD.89.123501}
}

@article{EFDE11,
title = {Inflation in entropic cosmology: Primordial perturbations and non-Gaussianities},
journal = {Physics Letters B},
volume = {697},
number = {4},
pages = {280-287},
year = {2011},
issn = {0370-2693},
doi = {https://doi.org/10.1016/j.physletb.2011.02.020},
author = {Yi-Fu Cai and Emmanuel N. Saridakis}
}

@article{EFDE12,
  title = {Generalizing the running vacuum energy model and comparing with the entropic-force models},
  author = {Basilakos, Spyros and Polarski, David and Sol\`a, Joan},
  journal = {Phys. Rev. D},
  volume = {86},
  issue = {4},
  pages = {043010},
  numpages = {13},
  year = {2012},
  month = {Aug},
  publisher = {American Physical Society},
  doi = {10.1103/PhysRevD.86.043010},
  url = {https://link.aps.org/doi/10.1103/PhysRevD.86.043010}
}

@article{EFDE13,
  title = {Non-adiabatic-like accelerated expansion of the late universe in entropic cosmology},
  author = {Komatsu, Nobuyoshi and Kimura, Shigeo},
  journal = {Phys. Rev. D},
  volume = {87},
  issue = {4},
  pages = {043531},
  numpages = {15},
  year = {2013},
  month = {Feb},
  publisher = {American Physical Society},
  doi = {10.1103/PhysRevD.87.043531},
  url = {https://link.aps.org/doi/10.1103/PhysRevD.87.043531}
}

@article{DESI1,
doi = {10.1088/1475-7516/2025/02/021},
url = {https://dx.doi.org/10.1088/1475-7516/2025/02/021},
year = {2025},
month = {feb},
publisher = {IOP Publishing},
volume = {2025},
number = {02},
pages = {021},
author = {The DESI collaboration and Adame, A.G. and Aguilar, J. and Ahlen, S. and Alam, S. and et al},
title = {DESI 2024 VI:  cosmological constraints from the measurements of baryon acoustic oscillations},
journal = {Journal of Cosmology and Astroparticle Physics}
}

@article{DESI2,
  title = {DESI DR2 results. II. Measurements of baryon acoustic oscillations and cosmological constraints},
  author = {Abdul Karim, M. and Aguilar, J. and Ahlen, S. and Alam, S. and et al},
  collaboration = {DESI Collaboration},
  journal = {Phys. Rev. D},
  volume = {112},
  issue = {8},
  pages = {083515},
  numpages = {40},
  year = {2025},
  month = {Oct},
  publisher = {American Physical Society},
  doi = {10.1103/tr6y-kpc6},
  url = {https://link.aps.org/doi/10.1103/tr6y-kpc6}
}

@article{DESI3,
  title = {Extended dark energy analysis using DESI DR2 BAO measurements},
  author = {Lodha, K. and Calderon, R. and Matthewson, W. L. and Shafieloo, A. and et al},
  collaboration = {DESI Collaboration},
  journal = {Phys. Rev. D},
  volume = {112},
  issue = {8},
  pages = {083511},
  numpages = {27},
  year = {2025},
  month = {Oct},
  publisher = {American Physical Society},
  doi = {10.1103/w4c6-1r5j},
  url = {https://link.aps.org/doi/10.1103/w4c6-1r5j}
}

@article{DESI4,
  title = {Constraints on neutrino physics from DESI DR2 BAO and DR1 full shape},
  author = {Elbers, W. and Aviles, A. and Noriega, H. E. and Chebat, D. and et al},
  collaboration = {DESI Collaboration},
  journal = {Phys. Rev. D},
  volume = {112},
  issue = {8},
  pages = {083513},
  numpages = {33},
  year = {2025},
  month = {Oct},
  publisher = {American Physical Society},
  doi = {10.1103/w9pk-xsk7},
  url = {https://link.aps.org/doi/10.1103/w9pk-xsk7}
}

@ARTICLE{DESI5,
       author = {{Gu}, Gan and {Wang}, Xiaoma and {Wang}, Yuting and {Zhao}, Gong-Bo and et al},
        title = "{Dynamical Dark Energy in light of the DESI DR2 Baryonic Acoustic Oscillations Measurements}",
      journal = {arXiv e-prints},
         year = 2025,
        month = apr,
          eid = {arXiv:2504.06118},
        pages = {arXiv:2504.06118},
          doi = {10.48550/arXiv.2504.06118},
archivePrefix = {arXiv},
       eprint = {2504.06118},
 primaryClass = {astro-ph.CO},
       adsurl = {https://ui.adsabs.harvard.edu/abs/2025arXiv250406118G}
}

@article{latest-EFDE1,
title = {Observational constraints on entropic cosmology},
journal = {Physics Letters B},
volume = {864},
pages = {139415},
year = {2025},
issn = {0370-2693},
doi = {https://doi.org/10.1016/j.physletb.2025.139415},
author = {Javier Chagoya and I. Díaz-Saldaña and Mario H. Amante and J.C. López-Domínguez and M. Sabido}
}

@article{latest-EFDE2,
  title = {Mass-to-horizon entropic cosmology: A unified thermodynamic pathway to cosmic acceleration},
  author = {Denkiewicz, Tomasz and Gohar, Hussain},
  journal = {Phys. Rev. D},
  volume = {113},
  issue = {6},
  pages = {063564},
  numpages = {15},
  year = {2026},
  month = {Mar},
  publisher = {American Physical Society},
  doi = {10.1103/zqkb-bqth},
  url = {https://link.aps.org/doi/10.1103/zqkb-bqth}
}

@article{latest-EFDE3,
title = {Multicomponent entropic cosmology model with generalized entropy},
journal = {Physics of the Dark Universe},
volume = {45},
pages = {101504},
year = {2024},
issn = {2212-6864},
doi = {https://doi.org/10.1016/j.dark.2024.101504},
author = {Facundo Abaca and Dario Javier Zamora}
}

@article{latest-EFDE4,
title = {On the foundations of entropic cosmologies: Inconsistencies, possible solutions and dead end signs},
journal = {Physics Letters B},
volume = {855},
pages = {138781},
year = {2024},
issn = {0370-2693},
doi = {https://doi.org/10.1016/j.physletb.2024.138781},
author = {Hussain Gohar and Vincenzo Salzano}
}

@article{latest-EFDE5,
    author = "Zamora, D. J. and Tsallis, C.",
    title = "{Thermodynamically consistent entropic late-time cosmological acceleration}",
    eprint = "2201.03385",
    archivePrefix = "arXiv",
    primaryClass = "gr-qc",
    doi = "10.1140/epjc/s10052-022-10645-x",
    journal = "Eur. Phys. J. C",
    volume = "82",
    number = "8",
    pages = "689",
    year = "2022"
}

@article{latest-EFDE6,
title = {Thermodynamically consistent entropic-force cosmology},
journal = {Physics Letters B},
volume = {827},
pages = {136967},
year = {2022},
issn = {0370-2693},
doi = {https://doi.org/10.1016/j.physletb.2022.136967},
author = {D.J. Zamora and C. Tsallis}
}

@article{effective-matter1,
doi = {10.3847/1538-4357/836/1/43},
url = {https://doi.org/10.3847/1538-4357/836/1/43},
year = {2017},
month = {feb},
publisher = {The American Astronomical Society},
volume = {836},
number = {1},
pages = {43},
author = {Solà, Joan and Gómez-Valent, Adrià and Pérez, Javier de Cruz},
title = {First Evidence of Running Cosmic Vacuum: Challenging the Concordance Model},
journal = {The Astrophysical Journal}
}

@misc{TCC1,
      title={Why the DESI Results Should Not Be A Surprise}, 
      author={Robert Brandenberger},
      year={2025},
      eprint={2503.17659},
      archivePrefix={arXiv},
      primaryClass={astro-ph.CO},
      url={https://arxiv.org/abs/2503.17659}, 
}

@article{TCC2,
  title = {Trans-Planckian problem of inflationary cosmology},
  author = {Martin, J\'er\^ome and Brandenberger, Robert H.},
  journal = {Phys. Rev. D},
  volume = {63},
  issue = {12},
  pages = {123501},
  numpages = {16},
  year = {2001},
  month = {May},
  publisher = {American Physical Society},
  doi = {10.1103/PhysRevD.63.123501},
  url = {https://link.aps.org/doi/10.1103/PhysRevD.63.123501}
}

@article{TCC3,
    author = "Bedroya, Alek and Vafa, Cumrun",
    title = "{Trans-Planckian Censorship and the Swampland}",
    eprint = "1909.11063",
    archivePrefix = "arXiv",
    primaryClass = "hep-th",
    doi = "10.1007/JHEP09(2020)123",
    journal = "JHEP",
    volume = "09",
    pages = "123",
    year = "2020"
}

@article{TCC4,
doi = {10.1088/1475-7516/2025/08/041},
url = {https://doi.org/10.1088/1475-7516/2025/08/041},
year = {2025},
month = {aug},
publisher = {IOP Publishing},
volume = {2025},
number = {08},
pages = {041},
author = {Li, Chunyu and Wang, Junkai and Zhang, Dongdong and Saridakis, Emmanuel N. and Cai, Yi-Fu},
title = {Quantum gravity meets DESI: dynamical dark energy in light of the trans-Planckian censorship conjecture},
journal = {Journal of Cosmology and Astroparticle Physics}
}

@article{quintom-B1,
title = {Dark energy constraints from the cosmic age and supernova},
journal = {Physics Letters B},
volume = {607},
number = {1},
pages = {35-41},
year = {2005},
issn = {0370-2693},
doi = {https://doi.org/10.1016/j.physletb.2004.12.071},
author = {Bo Feng and Xiulian Wang and Xinmin Zhang}
}

@article{quintom-B2,
  title = {Uncorrelated estimates of dark energy evolution},
  author = {Huterer, Dragan and Cooray, Asantha},
  journal = {Phys. Rev. D},
  volume = {71},
  issue = {2},
  pages = {023506},
  numpages = {5},
  year = {2005},
  month = {Jan},
  publisher = {American Physical Society},
  doi = {10.1103/PhysRevD.71.023506},
  url = {https://link.aps.org/doi/10.1103/PhysRevD.71.023506}
}

@ARTICLE{pantheonPLUS1,
       author = {{Scolnic}, Dan and {Brout}, Dillon and {Carr}, Anthony and {Riess}, Adam G. and et al},
        title = "{The Pantheon+ Analysis: The Full Data Set and Light-curve Release}",
      journal = {\apj},
         year = 2022,
        month = oct,
       volume = {938},
       number = {2},
          eid = {113},
        pages = {113},
          doi = {10.3847/1538-4357/ac8b7a},
archivePrefix = {arXiv},
       eprint = {2112.03863},
 primaryClass = {astro-ph.CO},
       adsurl = {https://ui.adsabs.harvard.edu/abs/2022ApJ...938..113S}
}

@article{pantheonPLUS2,
doi = {10.3847/1538-4357/ac8e04},
url = {https://dx.doi.org/10.3847/1538-4357/ac8e04},
year = {2022},
month = {oct},
publisher = {The American Astronomical Society},
volume = {938},
number = {2},
pages = {110},
author = {Brout, Dillon and Scolnic, Dan and Popovic, Brodie and Riess, Adam G. and et al},
title = {The Pantheon+ Analysis: Cosmological Constraints},
journal = {The Astrophysical Journal}
}

@article{Mb,
doi = {10.1088/0067-0049/192/1/1},
url = {https://doi.org/10.1088/0067-0049/192/1/1},
year = {2010},
month = {dec},
publisher = {The American Astronomical Society},
volume = {192},
number = {1},
pages = {1},
author = {Conley, A. and Guy, J. and Sullivan, M. and Regnault, N. and et al},
title = {SUPERNOVA CONSTRAINTS AND SYSTEMATIC UNCERTAINTIES FROM THE FIRST THREE YEARS OF THE SUPERNOVA LEGACY SURVEY*},
journal = {The Astrophysical Journal Supplement Series}
}

@article{CMB,
doi = {10.1088/1475-7516/2019/07/005},
url = {https://doi.org/10.1088/1475-7516/2019/07/005},
year = {2019},
month = {jul},
publisher = {},
volume = {2019},
number = {07},
pages = {005},
author = {Zhai, Zhongxu and Wang, Yun},
title = {Robust and model-independent cosmological constraints from distance measurements},
journal = {Journal of Cosmology and Astroparticle Physics}
}

@Article{MCMC1,
AUTHOR = {Padilla, Luis E. and Tellez, Luis O. and Escamilla, Luis A. and Vazquez, Jose Alberto},
TITLE = {Cosmological Parameter Inference with Bayesian Statistics},
JOURNAL = {Universe},
VOLUME = {7},
YEAR = {2021},
NUMBER = {7},
ARTICLE-NUMBER = {213},
URL = {https://www.mdpi.com/2218-1997/7/7/213},
ISSN = {2218-1997},
DOI = {10.3390/universe7070213}
}

@article{MCMC2,
  title = {Efficient sampling of fast and slow cosmological parameters},
  author = {Lewis, Antony},
  journal = {Phys. Rev. D},
  volume = {87},
  issue = {10},
  pages = {103529},
  numpages = {11},
  year = {2013},
  month = {May},
  publisher = {American Physical Society},
  doi = {10.1103/PhysRevD.87.103529},
  url = {https://link.aps.org/doi/10.1103/PhysRevD.87.103529}
}

@article{emcee,
doi = {10.1086/670067},
url = {https://dx.doi.org/10.1086/670067},
year = {2013},
month = {feb},
publisher = {University of Chicago Press},
volume = {125},
number = {925},
pages = {306},
author = {Foreman-Mackey, Daniel and Hogg, David W. and Lang, Dustin and Goodman, Jonathan},
title = {emcee: The MCMC Hammer},
journal = {Publications of the Astronomical Society of the Pacific}
}

@article{Verlinde2017,
	title = {Emergent Gravity and the Dark Universe},
	pages = {016},
	author = {Verlinde, Erik P.},
	journal = {SciPost Phys.},
	volume = {2},
	year = {2017},
	publisher = {SciPost},
	doi = {10.21468/SciPostPhys.2.3.016},
	url = {https://scipost.org/10.21468/SciPostPhys.2.3.016}
}

@article{Bekenstein1973,
  title = {Black Holes and Entropy},
  author = {Bekenstein, Jacob D.},
  journal = {Phys. Rev. D},
  volume = {7},
  issue = {8},
  pages = {2333--2346},
  numpages = {0},
  year = {1973},
  month = {Apr},
  publisher = {American Physical Society},
  doi = {10.1103/PhysRevD.7.2333},
  url = {https://link.aps.org/doi/10.1103/PhysRevD.7.2333}
}

@article{Hawking1974,
  author  = {Hawking, S. W.},
  title   = {Black hole explosions?},
  journal = {Nature},
  volume  = {248},
  number  = {5443},
  pages   = {30--31},
  year    = {1974},
  month   = {mar},
  doi     = {10.1038/248030a0},
  url     = {https://doi.org/10.1038/248030a0}
}

@article{KomatsuKimura2015,
  title = {Cosmic microwave background radiation temperature in a dissipative universe},
  author = {Komatsu, Nobuyoshi and Kimura, Shigeo},
  journal = {Phys. Rev. D},
  volume = {92},
  issue = {4},
  pages = {043507},
  numpages = {12},
  year = {2015},
  month = {Aug},
  publisher = {American Physical Society},
  doi = {10.1103/PhysRevD.92.043507},
  url = {https://link.aps.org/doi/10.1103/PhysRevD.92.043507}
}

@article{York1972,
  title = {Role of Conformal Three-Geometry in the Dynamics of Gravitation},
  author = {York, James W.},
  journal = {Phys. Rev. Lett.},
  volume = {28},
  issue = {16},
  pages = {1082--1085},
  numpages = {0},
  year = {1972},
  month = {Apr},
  publisher = {American Physical Society},
  doi = {10.1103/PhysRevLett.28.1082},
  url = {https://link.aps.org/doi/10.1103/PhysRevLett.28.1082}
}

@article{GibbonsHawking1977,
  title = {Action integrals and partition functions in quantum gravity},
  author = {Gibbons, G. W. and Hawking, S. W.},
  journal = {Phys. Rev. D},
  volume = {15},
  issue = {10},
  pages = {2752--2756},
  numpages = {0},
  year = {1977},
  month = {May},
  publisher = {American Physical Society},
  doi = {10.1103/PhysRevD.15.2752},
  url = {https://link.aps.org/doi/10.1103/PhysRevD.15.2752}
}

@article{GarciaBellido2024a,
title = {Cosmic entropic acceleration from supermassive black hole growth},
journal = {Physics of the Dark Universe},
volume = {44},
pages = {101491},
year = {2024},
issn = {2212-6864},
doi = {https://doi.org/10.1016/j.dark.2024.101491},
url = {https://www.sciencedirect.com/science/article/pii/S2212686424000736},
author = {Juan García-Bellido}
}

@article{GarciaBellido2024b,
title = {Dark Energy predictions from GREA: Background and linear perturbation theory},
journal = {Physics of the Dark Universe},
volume = {45},
pages = {101533},
year = {2024},
issn = {2212-6864},
doi = {https://doi.org/10.1016/j.dark.2024.101533},
url = {https://www.sciencedirect.com/science/article/pii/S2212686424001158},
author = {Juan García-Bellido}
}

@article{KaulMajumdar2000,
  title = {Logarithmic Correction to the Bekenstein-Hawking Entropy},
  author = {Kaul, Romesh K. and Majumdar, Parthasarathi},
  journal = {Phys. Rev. Lett.},
  volume = {84},
  issue = {23},
  pages = {5255--5257},
  numpages = {0},
  year = {2000},
  month = {Jun},
  publisher = {American Physical Society},
  doi = {10.1103/PhysRevLett.84.5255},
  url = {https://link.aps.org/doi/10.1103/PhysRevLett.84.5255}
}

@article{DasMajumdarBhaduri2002,
doi = {10.1088/0264-9381/19/9/302},
url = {https://doi.org/10.1088/0264-9381/19/9/302},
year = {2002},
month = {apr},
publisher = {},
volume = {19},
number = {9},
pages = {2355},
author = {Saurya Das and Parthasarathi Majumdar and Rajat K Bhaduri},
title = {General logarithmic corrections to black-hole entropy},
journal = {Classical and Quantum Gravity}
}

\end{document}